\documentclass[pre,aps,twocolumn,floatfix]{revtex4-1}

\usepackage{amssymb,amsmath,calc,ifsym,bbding,wasysym,amsfonts}

\pdfoutput=1
\usepackage{grfext}
\usepackage{grffile} 
\usepackage{xspace}
\usepackage{pifont,txfonts}

\usepackage{xpatch}
\makeatletter
\xpatchcmd{\@ssect@ltx}{\@xsect}{\protected@edef\@currentlabelname{#8}\@xsect}{}{}
\xpatchcmd{\@sect@ltx}{\@xsect}{\protected@edef\@currentlabelname{#8}\@xsect}{}{}
\makeatother
\usepackage{hyperref}

\usepackage[usenames]{color}
\usepackage{bm}
\usepackage[normalem]{ulem}
\usepackage{graphicx}

\PrependGraphicsExtensions*{.pdf,.PDF}  
\newcommand{\Revision}[1]{\textcolor{black}{#1}}

\newcommand{\eg}{\textit{e.g.}\xspace}
\newcommand{\etal}{\textit{et al.}\xspace}

\newcommand{\ecc}{\varepsilon_\text{CC}}
\newcommand{\esc}{\varepsilon_\text{SC}}
\newcommand{\ess}{\varepsilon_\text{HH}}
\newcommand{\ehh}{\varepsilon_\text{HH}}
\newcommand{\eph}{\varepsilon_\text{PH}}
\newcommand{\ephr}{\varepsilon_\text{PH}/\varepsilon_\text{HH}}

\newcommand{\eangle}{\varepsilon_\text{angle}}
\newcommand{\kshell}{\kappa_\text{s}}

\newcommand{\kt}{k_\text{B}T}

\newcommand{\rb}{r_\text{b}}

\newcommand{\sub}[2]{ _{\mathrm{#1}#2}}

\newcommand{\rcut}{r_\text{cut}}
\newcommand{\du}{\sigma_0}
\newcommand{\surfTension}{\tau}

\newcommand{\rhopT}{\rho_\text{p}^\text{T}}
\newcommand{\rhohT}{\rho_\text{h}^\text{T}}
\newcommand{\rhocT}{\rho_\text{c}^\text{T}}

\newcommand{\rhop}{\rho_\text{p}}
\newcommand{\rhoh}{\rho_\text{h}}
\newcommand{\rhoc}{\rho_\text{c}}
\newcommand{\rhocbar}{\bar{\rho_\text{c}}}
\newcommand{\rhoStar}{\rho^{*}}

\newcommand{\muc}{\mu_\text{c}}
\newcommand{\mup}{\mu_\text{p}}
\newcommand{\muh}{\mu_\text{h}}
\newcommand{\mucliq}{\mu_\text{c}^\text{liq}}
\newcommand{\dmu}{\Delta \mu_\text{h}}

\newcommand{\dmuc}{\Delta \mu_\text{c}}
\newcommand{\dmuprime}{{\Delta \mu_\text{h}'}}

\newcommand{\Nmin}{n_\text{min}}
\newcommand{\nc}{n_\text{c}}
\newcommand{\nh}{n_\text{h}}
\newcommand{\np}{n_\text{p}}

\newcommand{\Eelastic}{E_\text{elastic}}
\newcommand{\Ehelfrich}{E_\text{bend}}
\newcommand{\Ebend}{U_\text{bend}}
\newcommand{\Gshell}{U_\text{shell}}

\newcommand{\gh}{g_\text{hh}}
\newcommand{\gp}{g_\text{ph}}
\newcommand{\ghc}{g_\text{hc}}
\newcommand{\gpc}{g_\text{pc}}

\newcommand{\dgWrap}{\Delta \Omega_\text{wrap}}

\newcommand{\Rsp}{R_\text{0}}

\newcommand{\gammaSp}{\gamma_\text{0}}

\newcommand{\nb}{n_\text{B}}
\newcommand{\gammab}{\gamma_\text{B}}

\newcommand{\dgpent}{\Delta G_\text{p}}

\newcommand{\nStar}{{n^{*}}}

\begin{document}

\title{The role of the encapsulated cargo in microcompartment assembly}
\author{Farzaneh Mohajerani}
\affiliation{Martin Fisher School of Physics, Brandeis University, Waltham, MA, USA.}
\author{Michael F. Hagan}
\affiliation{Martin Fisher School of Physics, Brandeis University, Waltham, MA, USA.}

\begin{abstract}
Bacterial microcompartments are large, roughly icosahedral shells that assemble around enzymes and reactants involved in certain metabolic pathways in bacteria. Motivated by microcompartment assembly, we use coarse-grained computational and theoretical modeling to study the factors that control the size and morphology of a protein shell assembling around hundreds to thousands of molecules.  We perform dynamical simulations of shell assembly in the presence and absence of cargo over a range \Revision{of} interaction strengths, subunit and cargo stoichiometries, and the shell spontaneous curvature. Depending on these parameters, we find that the presence of a cargo can either increase or decrease the size of a shell relative to its intrinsic spontaneous curvature, as seen in recent experiments. These features are controlled by a balance of kinetic and thermodynamic effects, and the shell size is assembly pathway dependent. We discuss implications of these results for synthetic biology efforts to target new enzymes to microcompartment interiors.
\end{abstract}

\maketitle


While it has long been recognized that membrane-bound organelles organize the cytoplasm of eukaryotes, it is now evident that protein-based compartments play a similar role in many organisms. For example, bacterial microcompartments (BMCs) are icosahedral proteinaceous organelles that assemble around enzymes and reactants to compartmentalize certain metabolic pathways \cite{Kerfeld2010,Axen2014,Shively1998,Bobik1999,Erbilgin2014,Petit2013,Price1991,Shively1973,Shively1973a,Kerfeld2015}. BMCs are found in at least 20\% of bacterial species \cite{Bobik2006a,AbdulRahman2013,Axen2014}, where they enable functions such as growth, pathogenesis, and carbon fixation \cite{Chowdhury2014,Kerfeld2016,Kerfeld2015,Polka2016,Bobik2015,Kerfeld2010}. Other protein shells act as compartments in bacteria and archea, such as encapsulins \cite{Sutter2008} and gas vesicles \cite{Pfeifer2012,Sutter2008}, and even in eukaryotes (\eg vault particles \cite{Kickhoefer1998}). Understanding the factors that control the assembly of BMCs and other protein-based organelles is a fundamental aspect of cell biology. From a synthetic biology perspective, understanding factors that control packaging of the interior cargo will allow reengineering BMCs as nanocompartments that encapsulate a programmable set of enzymes, to introduce new or improved metabolic pathways into bacteria or other organisms (\eg \cite{Kerfeld2015,Bonacci2012,Parsons2010,Choudhary2012,Lassila2014,SliningerLee2017,SliningerLee2017a,Quin2016,Chessher2015,Cai2016,Huttanus2017})]. More broadly, understanding how the properties of a cargo affect the assembly of its encapsulating container is important for drug delivery and nanomaterials applications.

Despite atomic resolution structures of BMC shell proteins \cite{Tanaka2008,Kerfeld2010,Kerfeld2015,Sutter2017}, the factors that control the size and morphology of assembled shells remain incompletely understood. BMCs are large and polydisperse (40-600 nm diameter), with a roughly icosahedral protein shell surrounding up to thousands of copies of enzymes \cite{Price1991,Shively1973,Shively1973a,Iancu2007,Iancu2010,Kerfeld2010,Tanaka2008}. For example, the best studied BMC is the carboxysome, which encapsulates RuBisCO and carbonic anhydrase to facilitate carbon fixation in cyanobacteria \cite{Kerfeld2010,Schmid2006,Iancu2007,Tanaka2008}. BMC shells assemble from \Revision{multiple paralogous protein species, which respectively form homo-pentameric, homo-hexameric, and pseudo-hexameric (homo-trimeric) oligomers \cite{Tanaka2008,Kerfeld2010,Sutter2017}}. Sutter \etal \cite{Sutter2017} recently obtained an atomic-resolution structure of a complete BMC shell in a recombinant system that assembles small (40 nm) empty shells (containing no cargo).  The structure follows the geometric principles of icosahedral virus capsids, exhibiting $T{=}9$ icosahedral symmetry in the Caspar-Klug nomenclature \cite{Caspar1962,Johnson1997} (meaning there are 9 proteins in the asymmetric unit). The pentamers, hexamers, and pseudo-hexamers occupy different local symmetry environments.

Although the Sutter \etal \cite{Sutter2017} structure marks a major advance in understanding microcompartment architectures, it is uncertain how this construction principle extends to  natural microcompartments, which are large (100-600 nm), polydisperse, and lack perfect icosahedral symmetry. Moreover, the effect of cargo on BMC shell size is hard to interpret from experiments. In some BMC systems, empty shells are smaller and more monodisperse than full shells \cite{Lassila2014,Sutter2017,Cai2016,Mayer2016}, whereas in other systems empty shells are larger than full ones \cite{Lehman2017}. Thus, the cargo may increase or decrease shell size.
 
The encapsulated cargo can also affect BMC assembly pathways. Microscopy experiments showed that $\beta$-carboxysomes (which encapsulate form 1B RuBisCO) undergo two-step assembly: first the enzymes coalesce into a `procarboxysome', then   shells assemble on and bud from the procarboxysome \cite{Cameron2013,Chen2013}. In contrast, electron micrographs suggest that $\alpha$-carboxysomes (another type of carboxysome that encapsulates form 1A RuBisCO) assemble in one step, with simultaneous shell assembly and cargo coalescence \cite{Iancu2010,Cai2015}. Our recent computational study \cite{Perlmutter2016} suggested that the assembly pathway depends on the affinity between cargo molecules.  However, that study was restricted to a single shell size, and thus could not investigate correlations between assembly pathway and shell size.

Numerous modeling studies have identified factors controlling the thermodynamic stability \cite{Chen2007,Bruinsma2003,Zandi2004} or dynamical formation \cite{Berger1994,Schwartz2000,Rapaport2004,Nguyen2009,Elrad2008,Johnston2010,Rapaport2010,Nguyen2008,Wagner2015a} of empty icosahedral shells with different sizes. For example, Wagner and Zandi showed that icosahedral shells can form when subunits sequentially and irreversibly add to a growing shell at positions which globally minimize the elastic energy, with the preferred shell size determined by the interplay of elastic moduli and protein spontaneous curvature. Several studies have also investigated the effect of templating by an encapsulated nanoparticle or RNA molecule on preferred shell size \cite{Elrad2008,Fejer2010,Kusters2015,Zandi2009}. However, the many-molecule cargo of a microcompartment is topologically different from a nucleic acid or nanoparticle, and does not template for a specific curvature or shell size.

Rotskoff and Geissler recently proposed that microcompartment size is determined by kinetic effects arising from templating by the cargo \cite{Rotskoff2017}. Using an elegant Monte Carlo (MC) algorithm they showed that proteins without spontaneous curvature, which form polydisperse aggregates in the absence of cargo, can form kinetically trapped closed shells around a cargo globule.   However, there are reasons to question the universality of this mechanism for microcompartment size control. Firstly, several recombinant BMC systems form small, monodisperse empty shells \cite{Lassila2014,Sutter2017,Cai2016,Mayer2016}, suggesting that the shell proteins have a non-zero spontaneous curvature even without cargo templating. Secondly, when Cameron \etal \cite{Cameron2013} overexpressed RuBisCO to form `supersized' procarboxysomes, carboxysome shells encapsulated only part of the complex, suggesting that there is a maximum radius of curvature that can be accommodated by the shell proteins. Thirdly, the kinetic mechanism is restricted to systems in which rates of shell association vastly exceed cargo coalescence rates, a condition which may not apply in biological microcompartment systems. Thus, despite this and other recent simulation studies of microcompartments \cite{Perlmutter2016,Mahalik2016,Rotskoff2017},  the factors which control BMC size and amount of encapsulated cargo remain unclear.

In this article we use equilibrium calculations and Brownian dynamics (BD) simulations on a minimal model to identify the factors that control the size of a microcompartment shell.  Although computationally more expensive than the MC algorithm of Ref. \cite{Rotskoff2017}, BD better describes cooperative cargo-shell motions and thus allows for any type of assembly pathway.
Using this capability, we explore the effect of cargo on shell size and morphology over a range of parameters leading to one-step or two-step assembly pathways. To understand the interplay between shell curvature and cargo templating, we consider two limits of shell protein interaction geometries:  zero spontaneous curvature and high spontaneous curvature, which respectively form flat sheets or small icosahedral shells in the absence of cargo.
\begin{figure*}
\centering{\includegraphics[width=1\columnwidth]{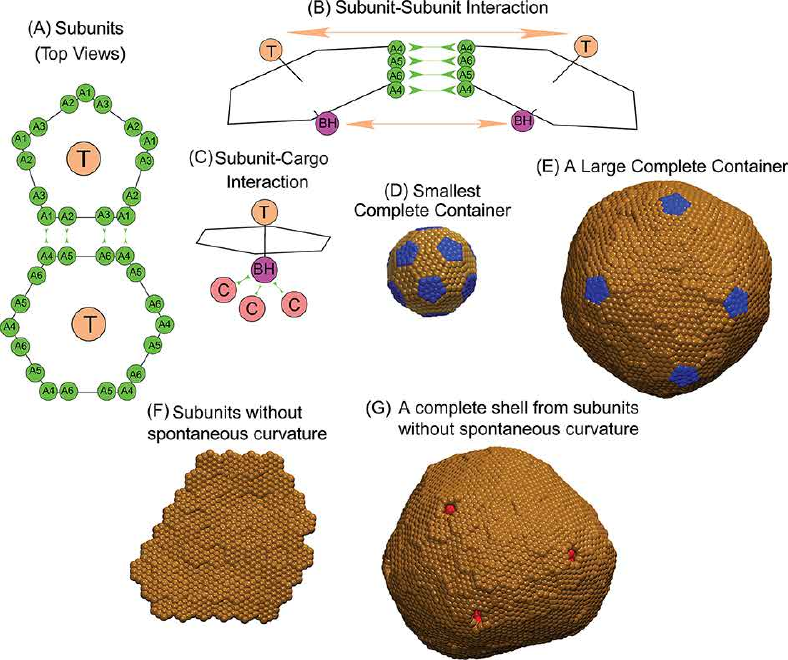}}
\caption{{ Description of the model.} \textbf{(A)} Each shell subunit contains `Attractors' (green circles) on the perimeter, a `Top' (tan circle, `T' ) in the center above the plane, and a `Bottom' (purple circle, `BH' and `BP' below the planes of the hexamer and the pentamer respectively).  \textbf{(B)} Interactions between Attractors drive subunit binding, while Top-Top and Bottom-Bottom repulsions control the subunit-subunit angle \Revision{and the shell bending modulus $\kshell$}. Attractions are indicated by green arrows in (A) for the pentamer-hexamer interface and in (B) for the hexamer-hexamer interface. \textbf{(C)} Only hexamer Bottom psuedoatoms `BH' bind cargo molecules (terra cotta circles, `C'). Excluder atoms (blue and brown pseudoatoms in \textbf{(D)}) placed in the plane of the `Top'  experience excluded volume interactions with the cargo. \textbf{(D)} The positions of excluder atoms in the preferred shell geometry for subunits with spontaneous curvature, a truncated icosahedron with 12 pentamers (blue) and 20 hexamers (brown). \textbf{(E)} Example of a shell that is larger than the preferred subunit geometry. \textbf{(F)} Subunits without spontaneous cuvature. \textbf{(G)} Example of hexamers without spontaneous curvature assembled around cargo (red).
\label{fig:Fig1-Schematic}
}
\end{figure*}
Our calculations find that the presence of cargo can increase or decrease shell size, depending on the stoichiometry of cargo and shell proteins, and the protein spontaneous curvature.  For shell proteins with high spontaneous curvature, we observe a strong correlation between assembly pathway and shell size, with two-step assembly leading to larger shells than single-step pathways or empty shell assembly. This result is consistent with the fact that $\beta$-carboxysomes tend to be larger than $\alpha$-carboxysomes. For shell proteins with zero spontaneous curvature, we find that introducing cargo can result in a well-defined shell size through several mechanisms, including the kinetic mechanism of Ref. \cite{Rotskoff2017} and the `finite-pool' effect due to a limited number of cargo particles available within the cell. However, spontaneous curvature of the shell proteins allows for robust shell formation over a wider range of parameter space.

\section{Methods}
\label{sec:methods}
\subsection*{Computational model}
\label{sec:comp_model}

\textbf{Shell subunits.}
BMC shells assemble from pentameric \Revision{(BMC-P)}, hexameric \Revision{(BMC-H)}, and pseudo-hexameric (trimeric, \Revision{BMC-T)} protein oligomers (\eg. Fig. 3A in Ref. \cite{Sutter2017} and Refs. \cite{Tanaka2008,Kerfeld2010,Kerfeld2015}). Experimental evidence suggests these oligomers are the basic assembly units, meaning that smaller complexes do not contribute significantly to the assembly process \cite{Sutter2016,Tanaka2008}. Although a recent atomic-resolution structure of synthetic BMC shells identifies specific roles for hexamers  and pseudo-hexamer  species \cite{Sutter2017}, it is unclear how these roles extend to larger shells. Therefore, for simplicity our model considers two basic assembly subunits, pentamers and hexamers, with the latter   fulfilling the roles of both hexamers and pseudo-hexamers.
We consider a minimal model which captures the directional interactions and excluded volume shape of subunits inferred from the recent structure \cite{Sutter2017}, and the fact that a closed shell is impermeable to cargo particles. Our model builds on previous models for virus assembly \cite{Perlmutter2013,Perlmutter2014,Perlmutter2015b, Wales2005, Fejer2009, Johnston2010,Ruiz-Herrero2015} and our recent model for the assembly around a fluid cargo \cite{Perlmutter2016}. However, while that model was specific to $T{=}3$ shells (containing 12 pentamers and 20 hexamers in a truncated icosahedron geometry), we have extended the model to describe shells of any size (see Fig.~\ref{fig:Fig1-Schematic}). A survey of other models which have been used for icosahedral shells can be found in Refs.~\cite{Hagan2014,Hagan2016,Mateu2013}.

\textbf{Shell-shell interactions.} Interactions between edges of BMC shell proteins are primarily driven by shape complementarity and hydrophobic interactions \cite{Sutter2017}. To mimic these short-ranged directionally specific interactions, each model subunit contains `Attractors' on its perimeter that mediate shell-shell attractions. Complementary Attractors on nearby subunits have short-range interactions (modeled by a Morse potential, Eq.~\eqref{eq:Morse} in \nameref{S1_ModelDetails}). Attractors which are not complementary do not interact. The arrangement of Attractors on subunit edges is shown in Fig.~\ref{fig:Fig1-Schematic}, with pairs of complementary Attractors indicated by green double-headed arrows. In the previous model \cite{Perlmutter2016} different hexamer edges interacted with either hexamers or pentamers, which made the model specific to the smallest possible shell, (a $T{=}3$ structure, Fig~\ref{fig:Fig1-Schematic}D). In this work, we allow for any shell geometry by making the hexamers six-fold symmetric, with each edge attracted to any edge on a nearby hexamer or pentamer. \Revision{However, because there is no experimental evidence of pentamer proteins (BMC-P) forming higher order assemblies (except non-specific aggregates) in the absence of hexamer proteins \cite{Keeling2013}, we do not consider attractive interactions between pairs of pentamers.}
The parameters $\ehh$ and $\eph$ scale the well-depths of the Morse potential between complementary Attractors for hexamer-hexamer and hexamer-pentamer interactions, and are thus the parameters that control the shell-shell binding affinity.
Further model details are in section~\nameref{S1_ModelDetails}.

To control the shell spontaneous curvature and bending modulus, each subunit contains a `Top' (type `TP' and `TH' for pentamers and hexamers respectively) pseudoatom above the plane of Attractors, and a `Bottom' pseudoatom (Types `BP' and `BH' for pentamers and hexamers respectively) below the Attractor plane. There are repulsive interactions (cutoff Lennard-Jones interactions, Eq.~\eqref{eq:LJ}) between Top-Top, Bottom-Bottom, and Top-Bottom pairs of pseudoatoms on nearby subunits. The relative sizes of the Top and Bottom pseudoatoms set the preferred subunit-subunit binding angle (and thus the spontaneous curvature), while the interaction strength (controlled by the well-depth parameter $\eangle$)  controls the shell bending modulus $\kshell$. We performed simulations of assembled shells to measure the relationship $\kshell(\eangle)$, as described in section \nameref{S1_ModelDetails}. The Top-Bottom interaction ensures that subunits do not bind in inverted orientations \cite{Johnston2010}. For subunits with no spontaneous curvature, we have extended simulations into the limit of unphysically small $\kshell$ values, for which the Top-Top and Bottom-Bottom repulsive interactions are insufficient to avoid partial subunit overlap. Therefore we have added an additional pseudoatom for subunits with no spontaneous curvature, a middle pseudoatom `M' placed in the center of the subunit in the plane of the attractors.  The addition of `M' pseudoatoms does not affect behaviors for $\eangle\ge 0.5$, and prevents overlaps below this range.

\textbf{Shell-cargo interactions.} Attractive interactions between hexamers and cargo are modeled by a a Morse potential with well-depth parameter $\esc$ between cargo particles (type `C')  and Bottom pseudoatoms on hexamers (type `BH'). These interactions represent shell-cargo attractions mediated by `encapsulation peptides' in BMCs  \cite{Kinney2012,Cameron2013,Fan2010,Aussignargues2015,Lehman2017}. \Revision{Because there is no experimental evidence that such encapsulation peptides interact with pentamers, in our model `BP' pseudoatoms do not interact} with cargo particles. 
We also add a layer of `Excluders'  in the plane of the `Top' pseudoatoms, which represent shell-cargo excluded volume interactions. Since the shell-shell interaction geometries are already controlled by the Attractor, Top, and Bottom pseudoatoms, we do not consider Excluder-Excluder interactions.

\textbf{Cargo.} \Revision{In carboxysome systems, attractions between RuBisCo particles are mediated by auxiliary proteins (\eg the protein CcmM in $\beta-$carboxysomes \cite{Cameron2013}). In refs \cite{Cameron2013,Chen2013} these interactions were shown to drive coalescence of RuBisCO prior to budding of $\beta-$carboxysomes assembled shells. Similarly, experiments and theory \cite{mackinder2016} support that protein-mediated phase separation of RuBisCO occurs in the pyrenoid, a dense complex of RuBisCO responsible for carbon fixing in plants. Since the complete phase diagram of RuBisCO and its auxiliary proteins is not known, we capture the possibility of cargo phase separation in the simplest manner possible by representing the cargo as spherical particles that interact via an attractive Lennard-Jones (LJ) potential, with well-depth $\ecc$. }

Perlmutter \etal \cite{Perlmutter2016} found that a more realistic, anisotropic model of the RuBisCO octomer holoenzyme did not qualitatively change assembly behaviors in comparison to spherical cargo particles \cite{Perlmutter2016}.

\Revision{The phase diagram of LJ particles contains regions of vapor, liquid, and solid and coexistence regimes \cite{lin2003}.  In this work we consider only one cargo density $0.0095/\du^3$, for which the vapor-liquid coexistence begins at $\ecc=1.5$ and the liquid-solid transition occurs at $\ecc=2.2$. Note that vapor-liquid coexistence in our finite system requires slightly stronger interactions than in the thermodynamic limit.}

\Revision{This model captures the excluded volume shape of subunits and their general binding modes observed in the microcompartment shell crystal structure \cite{Sutter2017}. Further refinements of the model are possible based on that structure, including an explicit representation of pseudo-hexamers and incorporating different preferred binding angles for pentamer-hexamer, hexamer-hexamer and hexamer-pseudo-hexamer interactions. It would be interesting to consider continued input of cargo or shell subunits into the system during assembly. Theoretical studies have suggested that a dynamical supply of subunits can affect the behavior of capsid assembly \cite{Boettcher2015,Dykeman2014,Castelnovo2014,Zhdanov2015,Hagan2011}.
}

\subsection*{Simulations}
We simulated assembly dynamics using the Langevin dynamics algorithm in HOOMD (which uses GPUs to efficiently simulate dynamics  \cite{Anderson2008}), and periodic boundary conditions to represent a bulk system. The subunits are modeled as rigid bodies \cite{Nguyen2011}.  Each simulation was performed in the NVT ensemble, using a set of fundamental units \cite{HoomdUnits} with $1\du$ defined as the circumradius of the pentagonal subunit (the cargo diameter is also set to 1 $\du$), and energies given in units of the thermal energy, $\kt$. 
The simulation time step was $0.005$ in dimensionless time units, and we performed $3 \times 10^{6}$ timesteps in each simulation unless mentioned otherwise.

\textit{Initial conditions.} We considered two types of initial conditions. Except where stated otherwise, simulations started from the `homogeneous' initial condition, in which subunits and (if present) cargo were initialized with random positions and orientations, excluding high-energy overlaps. In the `pre-equilibrated' initial condition, we first initialized cargo particles with random positions (excluding high-energy overlaps), and performed $10^5$ simulation timesteps to equilibrate the cargo particles. Shell subunits were then added to the simulation box with random positions and orientations, excluding high-energy overlaps.

\textit{Systems.} We simulated several systems as follows. For shell subunits with spontaneous curvature we set pentamer-hexamer and hexamer-hexamer angles consistent with the $T{=}3$ geometry (see Estimating the shell bending modulus in section \nameref{S2_Thermodynamics} ), and we set $\eangle=0.5$. \Revision{We first performed a set of empty-shell assembly simulations, with 360 hexamers, and varying number of pentamers, in a cubic box with side length $60\du$, with $\ehh=2.6\kt$ (the smallest interaction strength for which nucleation occurred). These simulations were performed for $10^7$ timesteps to obtain sufficient statistics at low pentamer concentrations despite nucleation being rare.}

For cargo encapsulation by subunits with spontaneous curvature, we simulated 2060 cargo particles, 180 pentamers, and 360 hexamers in a cubic box with side length $60\du$. Other parameters were the same as for the empty-shell simulations, except that we varied $\eph$, $\esc$, and $\esc$ as described in the main text.  All simulations with spontaneous curvature used $\eph\ge1.3 \ehh$ to ensure that the \Revision{shells with the $T{=}3$ geometry (or asymmetric shells with similar sizes)} were favored in the absence of cargo. We note that our results generalize to other ranges of shell interaction parameters, but this choice distinguishes effects due to cargo from those due to changes in the inherent preferred shell geometry.  \Revision{Simulations with strong cargo-cargo and cargo-shell interactions \Revision{($\ecc\ge 1.55$ and $\esc\ < 8.75$)} required a long timescale for pentamers to fill pentameric vacancies in the hexamer shell (discussed in Results). To observe pentamer adsorption, these simulations were run for up to $9 \times 10^{6}$ simulation timesteps.}

For simulations of `flat' subunits (with no spontaneous curvature), we considered a range of system sizes at fixed steady state cargo chemical potential,  with the number of cargo particles varying from 409 to 3275, and the box side length varying from $35\du$ to $70\du$. Since these were NVT simulations, we ensured that the final hexamer chemical potential was the same at each system size by setting the number of hexamers so that the concentration of free hexamers remaining after assembly of a complete shell was constant ($10^{-3}$ subunits/$\du^3$). The resulting number of hexamers varied from 109 to 581 in boxes with side lengths $35\du$ to $70\du$. The assembly outcomes were unchanged if instead we kept the total hexamer subunit concentration the same across all simulations. For each of these system sizes we performed simulations over a range of $\eangle$ to identify the maximum value of $\kshell$ at which assembly of a complete shell could occur. Simulations were stopped upon completion of a shell or after the maximum simulation time $t_\text{max}$ with $t_\text{max}=3\times 10^6$ timesteps for boxes with side length $\le 55\du$ and $t_\text{max}=8\times 10^6$ for boxes with side length $\ge 55 \du$.  The maximum simulation time was increased for large system sizes because the minimum time required for assembly of a complete shell increases linearly with the shell size \cite{Hagan2010}.

To estimate the relationship between the shell bending modulus $\kshell$ and the parameter $\eangle$ we performed additional simulations, in which we measured the total interaction energy of completely assembled shells as a function of $\eangle$ (see `Estimating the shell bending modulus' in section \nameref{S2_Thermodynamics}).

\textit{Sample sizes.} For simulations of shells with spontaneous curvature, we performed a minimum of 10 independent trials at each parameter set. To enable satisfactory statistics on shell size and morphology for parameter sets that result in at most one complete shell in the simulation box \ref{fig:Fig3-WithC0-AvgSize}, we performed additional trials such that at least 10 complete shells were simulated.
For flat subunits (Fig.~\ref{fig:Fig1-Schematic} F, G), we identified the maximum $\eangle$ in which a complete shell forms at each system size as follows. We first performed independent simulations over a range of $\eangle$ values, separated by increments in $\eangle$ of 0.02 for systems with box side length $\le 55\du$, and increments of 0.05 for systems with side length $\ge 55\du$. We performed 10 independent trials at each value of $\eangle$. For the largest value of $\eangle$ at which at least one of these trials resulted in a complete shell, we then performed 10 additional trials to obtain a more accurate estimate of the shell bending modulus $\kshell$ at the maximum $\eangle$.

\begin{figure*}[ht!]
\centering{\includegraphics[width=1.1\columnwidth]{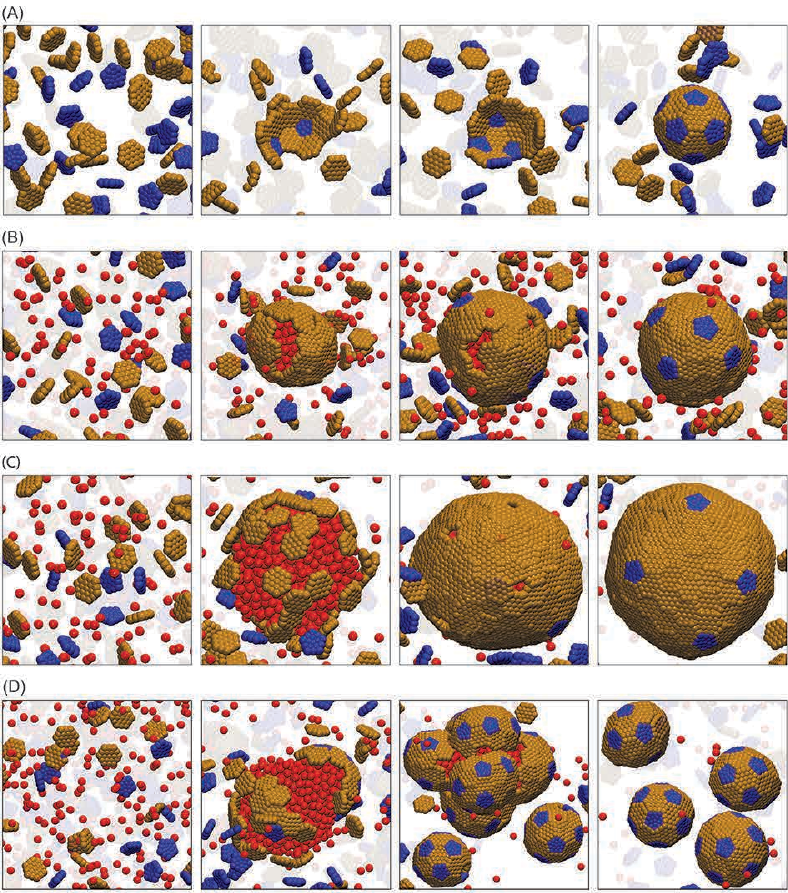}}
\caption{{ Snapshots from assembly trajectories of subunits with $T{=}3$ preferred curvature.}
\textbf{(A)}  Small $T{=}3$ shells (20 hexamers, 12 petamers) assembled without cargo at $\ess=2.6$ and pentamer/hexamer stoichiometric ratio $\rhop/\rhoh=0.5$. Notice that the intermediate in the third frame contains a hexamer where a pentamer is required for icosahedral symmetry. This hexamer eventually dissociates.
\textbf{(B)} One-step assembly with moderate cargo-cargo interaction strength, $\ecc=1.5$. A small nucleus of cargo and hexamer subunits forms, followed by simultaneous cargo coalescence, shell growth, and finally filling in of defects by pentamers subunits. The final structure has 68 hexamers, 12 pentamers, and 408 encapsulated cargo particles. Other parameters are hexamer-hexamer affinity $\ess$=1.8, ratio of pentamer/hexamer affinity $\eph/\ess=1.3$, and shell-cargo affinity $\esc=8.75$, and $\rhop/\rhoh=0.5$.
\textbf{(C)} Two-step assembly pathway for strong cargo-cargo affinity $\ecc=1.65$. Rapid cargo coalescence is followed by adsorption and assembly of shell subunits. The final structure has 167 hexamers, 12 pentamers, and 1520 encapsulated cargo particles. Other parameters are $\ess=1.8$, $\esc=8.5$, and $\rhop/\rhoh=0.5$.
\textbf{(D)} Assembly and budding of shells from a cargo globule, for high pentamer/hexamer affinity ratio $\eph/\ess=2.0$. Other parameters are $\ecc=1.65$ , $\ess=1.8$ , $\esc=8.5$ and $\rhop/\rhoh=0.8$. (We report energies in units of $\kt$ throughout this article.)
The shell bending modulus for all panels is $\kshell=10 \kt$.
\label{fig:Fig2-WithC0-Traj}
}
\end{figure*}
\begin{figure*}
\centering{\includegraphics[width=1.1\columnwidth]{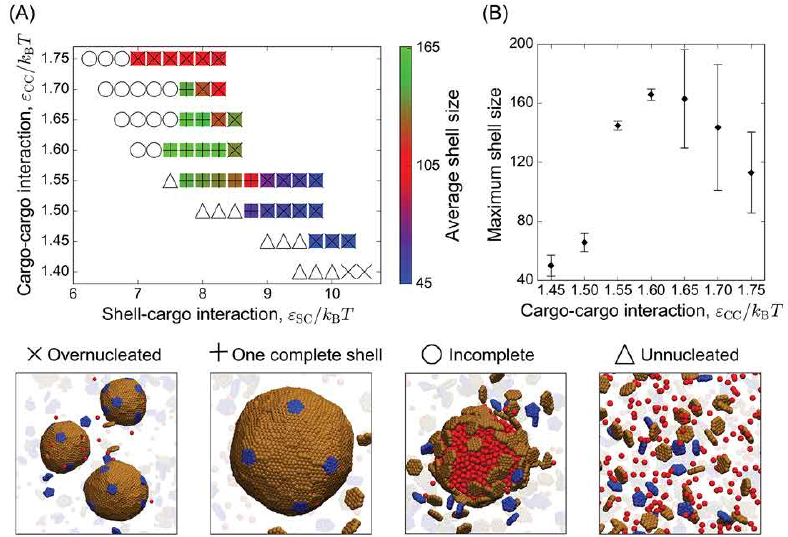}}
\caption{{ Dependence of the mean shell size and most probable morphology on the cargo-cargo and subunit-cargo affinities ($\ecc$ \& $\esc$).} \textbf{(A)} The mean shell size (number of hexamers + 12 pentamers) is indicated by the color bar, and the predominant morphology is indicated by symbols, with a snapshot corresponding to each morphology shown on the right. \textbf{(B)}  The mean shell size maximized over $\esc$ is shown as a function of $\ecc$. Other parameters in (A) and (B) are  $\ess=1.8$, $\rhop/\rhoh=0.5$, $\eph/\ess=1.3$, and $\kshell=10\kt$.
\label{fig:Fig3-WithC0-AvgSize}
}
\end{figure*}
\begin{figure}
\centering{\includegraphics[width=.95\columnwidth]{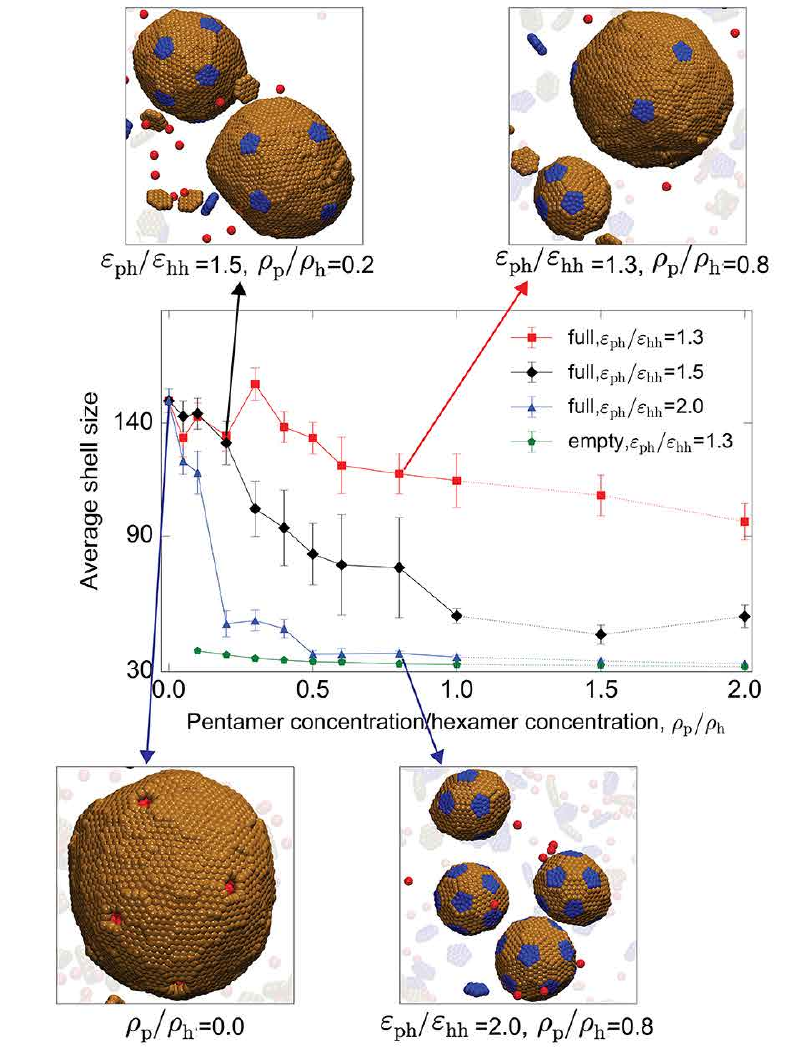}}
\caption{{ Dependence of shell size on the driving force for pentamer addition.} The mean shell size (number of hexamers + 12 pentamers) is shown as a function of the pentamer/hexamer stoichiometry ratio $\rhop/\rhoh$ for indicated values of the  pentamer/hexamer affinity ratio $\ephr$ for simulations with cargo. \Revision{Results from empty shell simulations are also shown for $\ephr=1.3$.} Snapshots of typical assembly morphologies for indicated parameter values are shown around the plot.
In these simulations the hexamer concentration, hexamer-hexamer affinity, and hexamer-shell affinity, and bending modulus were fixed at $\rhoh=1.7\times10^{-3}/ \du^{3}$, $\ess=1.8$, $\ecc=1.65$,  $\esc=8.5$, and $\kshell=10\kt$.
\label{fig:Fig4-WithC0-Pentamer}
}
\end{figure}

\section{Results and Discussion}
\label{sec:results}
To simulate the dynamics of microcompartment assembly, we build on the model developed by Perlmutter \etal \cite{Perlmutter2016}, which allowed only a single energy minimum shell geometry, corresponding to a $T{=}3$ icosahedral shell containing 12 pentamers and 20 hexamers. We have now extended the model to allow for closed shells of any size. Based on AFM experiments showing that BMC shell facets assemble from pre-formed hexamers \cite{Sutter2016}, and the fact that carboxysome major shell proteins crystallize as pentamers and hexamers \cite{Tanaka2008}, our model considers pentamers and hexamers as the basic assembly units. These are modeled as rigid bodies with short-range attractions along their edges, which drive hexamer-hexamer and hexamer-pentamer association. Repulsive subunit-subunit interactions control the preferred angle of subunit-subunit interactions, which sets the shell protein spontaneous curvature (Fig.~\ref{fig:Fig1-Schematic}A,B). To minimize the number of model parameters, we do not explicitly consider pseudo-hexamers; thus, the model hexamers play the role of both hexamers and pseudo-hexamers.

We particularly focus on carboxysomes, for which the most experimental evidence is available, although our model is sufficiently general that results are relevant to other microcompartment systems. In carboxysomes, interactions between the RuBisCO cargo and shell proteins are mediated by non-shell proteins containing `encapsulation peptides' \cite{Cameron2013,Kinney2012,Long2010,Long2007,Niederhuber2017,Cai2015,Rae2013}. For simplicity we model these interactions as direct-pair attractions between model cargo particles and shell subunits. Because there is no evidence that encapsulation peptides  interact with pentamers, in our model the cargo only interacts with hexamers.
Further details of the model and a thermodynamic analysis are given in section~\nameref{sec:comp_model} and section~\nameref{S2_Thermodynamics}.

There are numerous parameters which can affect shell size, including the interaction strengths among the various species of cargo and shell subunits, shell protein spontaneous curvature and bending modulus, and the concentration of each species. To facilitate interpretation of results from this vast parameter space, we focus our simulations on two extreme limits. In the first limit, we consider shell subunits with a spontaneous curvature that favors assembly of the smallest icosahedral shell, the $T{=}3$ structure with 12 pentamers and 20 hexamers (Fig.~\ref{fig:Fig1-Schematic}D). In the second limit we consider a system containing only hexamer subunits with no preferred curvature, which form flat sheets without cargo (Fig.~\ref{fig:Fig1-Schematic}F).
\begin{figure*}
\centering{\includegraphics[width=1.1\columnwidth]{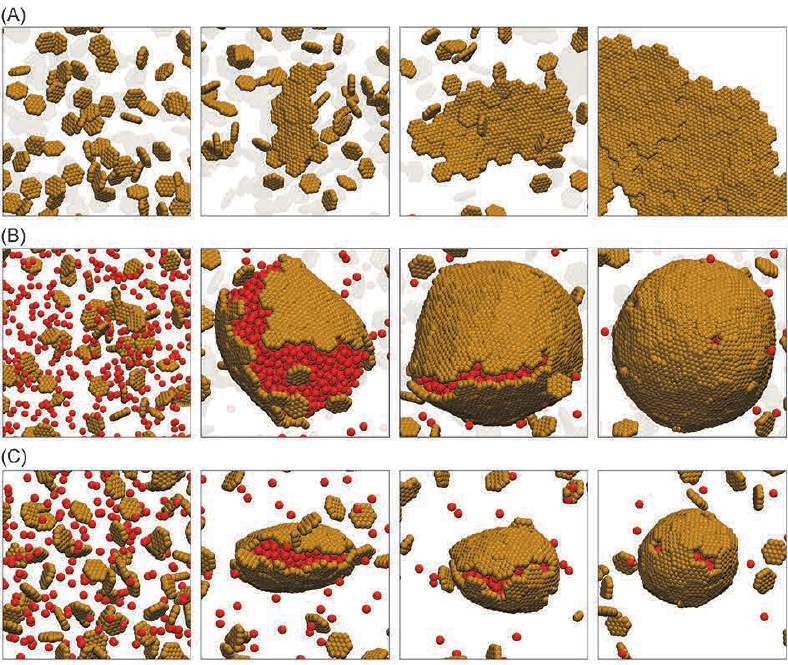}}
\caption{ { Snapshots of assembly trajectories for hexamer subunits with zero spontaneous curvature.} \textbf{(A)} Assembly with no cargo, for $\ess$=2.5, and shell bending modulus parameter $\eangle=0.1$ (shell bending modulus $\kshell\approx 20 \kt$).
\textbf{(B)} Assembly with cargo, for $\ess$=1.8, $\esc$=7.0, and $\eangle=0.08$ ($\kshell\approx 18 \kt$). The final shell has 231 and 2261 hexamers and cargo particles respectively, as well as 12 pentameric vacancies.
\textbf{(C)} Assembly with cargo in a small system with low shell bending modulus, for $\ess$=1.8, $\esc$=7.0, and $\eangle=0.015$ ($\kshell\approx 3 \kt$). The final shell has 71 and 361 hexamers and cargo particles respectively, 8 pentameric vacancies, and 2 double vacancies. An example of a double vacancy is visible in the front of the final frame.
\label{fig:Fig5-NoC0-Traj}
}
\end{figure*}
\subsection*{Cargo increases the size of shells with high spontaneous curvature}

We begin by considering shells with $T{=}3$ spontaneous curvature  (Fig.~\ref{fig:Fig1-Schematic}D). To isolate the effects of cargo on shell size, we consider shell-shell interaction parameters which favor pentamer insertion (setting the ratio of pentamer-hexamer and hexamer-hexamer affinities $\eph/\ess\ge1.3$) so that assembly without cargo results in \Revision{primarily $T{=}3$ empty shells for our ratio of pentamer to hexamer concentrations, $\rhop/\rhoh=0.5$, and results in shells close in size to the T=3 geometry at all of the stoichiometries we consider here.}

A typical assembly trajectory without cargo is shown in Fig.~\ref{fig:Fig2-WithC0-Traj} A. When simulating assembly around cargo, we set the hexamer-hexamer affinity $\ess \le 2.2$ (while maintaining $\eph/\ess\ge1.3$) so that assembly occurs only in the presence of cargo, and we vary cargo-cargo $\ecc$ and cargo-shell $\esc$ interaction strengths. Throughout this article, all energy values are given in units of the thermal energy, $\kt$.
\Revision{Except where mentioned otherwise, values of our simulation shell bending modulus $\kshell$ fall within the range estimated for $\beta-$carboxysomes from AFM nanoindention experiments $\kshell\in[1,25]\kt$ (see Ref. \cite{Faulkner2017} and section~\nameref{sec:paramValues} in \nameref{S2_Thermodynamics}); \Revision{simulations with shell spontaneous curvature use $\kshell=10-16\kt$.}}

\textbf{Assembly pathways.}
Consistent with previous simulations of $T{=}3$-specific shells \cite{Perlmutter2016}, we find that assembly proceeds by one-step and two-step pathways, with the type of pathway primarily determined by the strength of cargo-cargo interactions. For $\ecc\lesssim 1.5$ (Fig.~\ref{fig:Fig2-WithC0-Traj}B), the cargo lies at or below the border of phase coexistence, and there is a large barrier for cargo coalescence.  However, a fluctuation in the local density of hexamers allows nucleation of a small cargo globule and shell cluster, after which cargo condensation, shell subunit adsorption and assembly occur simultaneously. On the other hand, for $\ecc\gtrsim1.55\kt$ (Fig.~\ref{fig:Fig2-WithC0-Traj}C) a cargo globule coalesces rapidly. Hexamers then adsorb onto the cargo globule in a disordered manner, followed by reorganization and assembly. Since pentamers are not directly attracted to the cargo, they are mostly excluded for the pentamer/hexamer affinity ratio, $\eph/\ess=1.3$, considered in Fig.~\ref{fig:Fig2-WithC0-Traj}C.  However, the hexamers cannot form a closed surface around the globule since the spherical topology requires 12 five-fold defects \cite{Grason2016}. Interestingly, for moderate interaction strengths we find that shells satisfy this requirement by forming exactly 12 pentamer-sized vacancies in the shell, which are gradually filled in by pentamers.
Increasing the pentamer/hexamer affinity ratio to $\eph/\ess=2$ (Fig.~\ref{fig:Fig2-WithC0-Traj}D) allows pentamers to rapidly bind to adsorbed hexamers, creating additional shell curvature and thus driving the budding of small shells containing part of the globule in their interior.

\Revision{The shells assembled around cargo are larger and lack the perfect icosahedral symmetry of the intrinsic preferred shell geometry ($T{=}3$, 20 hexamers). Despite the lack of symmetry, most shells are closed, meaning that every hexamer and pentamer subunit interacts with respectively six and five neighboring subunits. The yield and fraction of complete shells are shown in Figs.~\ref{fig:FigS1-Yield} and \ref{fig:FigS2-Quality}. Once a complete shell forms with or without cargo, it is stable on assembly timescales even under infinite dilution of subunits. This hysteresis between assembly and disassembly is consistent with previous experimental and theoretical studies of virus assembly \cite{Singh2003,Roos2010,Uetrecht2010a,Hagan2006,Hagan2014}, and occurs because removal of the first subunit from a complete shell breaks multiple contacts thus incurring a large activation barrier.}

\Revision{Fig.~\ref{fig:FigS3-Symmetry} shows the Steinhardt icosahedral order parameter as a function of shell size along with snapshots of typical shells. We observe that the degree of icosahedral symmetry increases with shell size, and is correlated to the assembly pathway. Small shells that assemble by one-step pathways (with $\thicksim 50$ subunits) are clearly asymmetric, corresponding neither to icosahedral symmetry nor other symmetric low-energy minimum arrangements expected for shells in this size range \cite{Llorente2014}, whereas large shells  are nearly (though not perfectly) icosahedral. The lack of perfect symmetry likely arises because the hexamers form an elastic sheet, within which shell reorganization and defect diffusion are slow in comparison to assembly timescales. Based on analysis of assembly trajectories, we speculate that the higher degree of symmetry for large shells reflects the fact that pentamers are incorporated near the end of two-step pathways (filling in pentamer-sized vacancies) whereas pentamers incorporate early in one-step pathways. Because rearranging a pentamer within a shell requires breaking more bonds than does a vacancy, pentamer rearrangement is slower than vacancy diffusion.}

\textbf{Shell size depends on interaction strengths, subunit stoichiometry, and initial conditions.} Fig.~\ref{fig:Fig3-WithC0-AvgSize}A shows the mean size and predominant assembly morphology as a function of cargo-cargo and cargo-shell interaction strengths. Over a wide range of parameter space, shell sizes are larger than the $T{=}3$ size formed by empty shells (32 subunits), demonstrating that the cargo can robustly increase shell size. As the shell-cargo interaction is increased within the two-step regime ($\ecc\gtrsim1.55)$, there is a sequence of predominant assembly outcomes. Weak interactions lead to a disordered layer of shell subunits on the cargo globule, moderate interactions result in one complete shell, and overly strong interactions drive multiple nucleation events throughout the system. This over-nucleation decreases the mean shell size since the system becomes depleted of cargo and shell subunits. The one-step regime exhibits a similar sequence, except that instead of a disordered globule there is no nucleation for weak shell-cargo interactions.

\emph{Pathway dependence.} A striking feature of Figs.~\ref{fig:Fig2-WithC0-Traj} B and C is that the two-step assembly pathway leads to much larger shells than the one-step pathway, increasing the number of encapsulated cargo particles by more than a factor of five. We observe a similar correlation between shell size and assembly pathway across the range of simulated parameters. To emphasize the effect of cargo-cargo interactions on shell size, Fig.~\ref{fig:Fig3-WithC0-AvgSize}B shows the maximum shell size obtained as a function of $\ecc$ (maximized over $\esc$). We see a dramatic increase in shell size as the cargo-cargo interactions increase beyond $\ecc=1.5$, when the system transitions to two-step assembly pathways. The maximum shell size eventually decreases for $\ecc\gtrsim1.65$ due to over-nucleation.

\emph{Dependence on shell subunit stoichiometry.} To determine the effects of shell subunit stoichiometry, we performed simulations with varying concentrations $\rhop$ of pentamers subunits at fixed hexamer concentration. As shown in Fig.~\ref{fig:Fig4-WithC0-Pentamer}, increasing the pentamer concentration uniformly decreases the shell size. Since only 12 pentamers are required for a closed shell, increasing their chemical potential favors increased pentamer insertion and thus smaller total shell sizes. The effect depends on the pentamer-hexamer affinity; for the moderate pentamer-hexamer interactions considered above ($\eph=1.3\ess$), we observe a modest decrease in shell size of about 50\% with increasing pentamer concentration. In contrast, for strong pentamer-hexamer interactions ($\eph=2\ess$), even small concentrations of pentamers lead to rapid pentamer insertion and shells that are close in size to the minimum $T{=}3$ geometry. \Revision{At low pentamer stoichiometries we observe very large shells containing approximately 140 subunits; the shell size saturates because it is limited by the droplet size and multi-nucleation events that occur for these relatively strong cargo-cargo and cargo-shell interactions ($\ecc=1.65$ and $\esc=8.5$). In comparison, empty shells with $\rhop/\rhoh=0.1$ and $\eph=1.3$ have a mean size of $39$ subunits.}

\emph{Kinetics vs. thermodynamics.} \Revision{Our trajectories start from an out-of-equilibrium condition of unassembled subunits, and reorganization of complete shells it is slow in comparison to assembly timescales. Therefore the ensemble of shells that we observe in finite-time simulations can depend on both kinetic and thermodynamic effects. We performed several analyses to assess the relative importance of kinetics and thermodynamics.}

First, we investigated whether assembly morphologies depend on initial configurations. For notational clarity, we will refer to the initial condition for simulations described so far, in which shell subunits and cargo start from random positions, as the `homogeneous' initial condition. We performed a second set of simulations started from a `pre-equilibrated globule' initial condition, in which the cargo particles were allowed to completely phase separate before introduction of the shell subunits (see section \nameref{sec:methods}). When the cargo is below phase coexistence ($\ecc<1.5$ at the simulated cargo concentration) the two initial conditions produce identical results.

Above phase coexistence the pre-equilibrated globule leads to larger globule sizes in comparison to the homogeneous initial condition, since shell assembly tends to arrest globule coalescence. Correspondingly, the pre-equilibrated globule initial condition produces larger shells than  the homogeneous initial condition (Figs.~ \ref{fig:FigS4-PreEq-Hmgns-Traj}  and \ref{fig:FigS5-PreEq-Hmgns-AvgSize}). This effect is most significant at the boundary of phase coexistence ($\ecc\approx1.5$), since there is a large nucleation barrier to cargo coalescence.

This dependence on initial conditions demonstrates that kinetics quantitatively affects the size and morphology of assembled shells. However qualitative effects are limited by the degree of mismatch between the globule size and the shell preferred curvature; a large mismatch leads to budding of shells containing only part of the globule (Fig.~\ref{fig:FigS4-PreEq-Hmgns-Traj}).

To further evaluate whether assembly depends on kinetics or thermodynamics, we compared the dynamical simulation results against predictions of an equilibrium theory, based on rough estimates of equilibrium binding affinities and shell bending modulus values corresponding to our simulation parameters (section \nameref{S2_Thermodynamics}).  As shown in Figs.~\ref{fig:FigS6-Theory-WithC0-AvgSize} and \ref{fig:FigS7-WithC0-Pentamer-THeory}, the equilibrium dependence of the shell size on parameters exhibits similar qualitative trends as observed in the simulations, but the dynamical simulations exhibit larger variations in shell size than predicted at equilibrium.

\textbf{Mechanisms of size selection.}

By comparing results from the equilibrium model and simulation results from two sets of initial conditions, we determine that the effect of cargo on shell size arises from the competition of several effects.
The first two are equilibrium effects. Firstly, because only hexamers interact with the cargo, increasing the shell-cargo interaction increases the chemical potential of pentamers in the shell relative to hexamers. As noted above, decreasing pentamer adsorption favors larger shells, since there are only 12 pentamers in a complete shell (Fig.~\ref{fig:Fig3-WithC0-AvgSize} and \ref{fig:FigS6-Theory-WithC0-AvgSize} at low $\esc$). Similarly, decreasing the pentamer concentration $\rhop$ reduces pentamer insertion and thus increases shell size (Figs.~\ref{fig:Fig4-WithC0-Pentamer} and \ref{fig:FigS7-WithC0-Pentamer-THeory}). Secondly, however, increasing the shell-cargo interaction strength leads to a lower shell surface energy, which favors a larger surface-to-volume ratio and hence smaller shells. Above threshold values of $\ess$ and $\esc$, the second effect dominates (Figs.~\ref{fig:Fig3-WithC0-AvgSize} and \ref{fig:FigS6-Theory-WithC0-AvgSize} at high $\esc$). Due to these two competing effects, the equilibrium theory predicts a nonmonotonic dependence of the equilibrium shell size on $\esc$. \Revision{The equilibrium theory identifies other factors which affect the ratio of surface to bulk energy and thus shell size. For example, increasing the stoichiometric ratio of cargo to shell subunits decreases the cargo chemical potential thus favoring larger shells, consistent with a previous theoretical study on virus capsid assembly \cite{Zandi2009}.}

The tendency of the cargo to form spherical droplets also leads to kinetic effects on shell size, which depend on the relative rates of cargo coalescence and shell assembly. The sizes of the initial cargo globule and the final shell are correlated because the globule surface tension imposes a barrier to formation of shells with curvature radii that are smaller than the globule radius. Furthermore, since shell completion arrests globule coalescence, and stronger interactions drive faster assembly,  the final size of the globule and the shell decrease with increasing $\esc$ and $\ess$.  The assembly of larger shells in simulations started with the pre-equilibrated globule initial condition shows that this is at least partly a kinetic effect.

Finally, recall that above threshold values of $\ecc$ and $\esc$, interactions are sufficiently strong that nucleation occurs throughout the system. Once complete (small) shells assemble around these nascent droplets, subsequent coarsening of globule-shell complexes is arrested on relevant timescales, resulting in a broad, non-equilibrium distribution of shell sizes (Fig.~\ref{fig:Fig3-WithC0-AvgSize}B).

\begin{figure}
\centering{\includegraphics[width=1.\columnwidth]{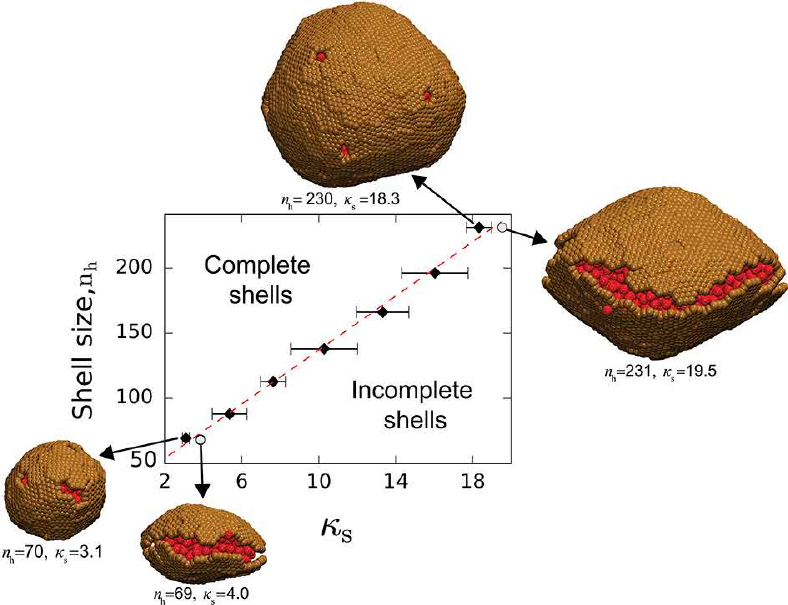}}
\caption{{ Size and morphology of shells assembled from subunits with no spontaneous curvature, for varying system sizes and  shell bending modulus $\kshell$.} The y-axis gives the number of subunits in the largest cluster at the final simulation frame. The $\blacklozenge$ symbols correspond to Brownian dynamics simulation results for the smallest system size in which a complete shell formed, and the dashed line shows the best fit of  Eq.~\eqref{eq:dgWrap} to this data. The snapshots show examples of the final morphology at indicated parameter values. Two snapshots are shown of shells just below the threshold size for completion, with corresponding parameters indicated by $\circ$ symbols. Other parameters are $\ecc=1.7$, $\ess=1.8$, and $\esc=7.0$.
}
\label{fig:Fig6-kShell-Size}
\end{figure}

\subsection*{Shell subunits with no spontaneous curvature.}
We now consider the opposite limit:  a system of `flat' hexamer subunits, which have zero spontaneous curvature and thus favor formation of flat sheets (Fig.~\ref{fig:Fig5-NoC0-Traj}A). Fig.~\ref{fig:Fig5-NoC0-Traj}B shows a typical assembly trajectory for flat subunits with $\ecc=1.7$, in which the cargo rapidly coalesces followed by adsorption and assembly of the hexamers. Interestingly, the shapes of assembly intermediates reflect the lack hexamer spontaneous curvature ---  hexamers initially assemble into flat sheet wrapped around the globule, deforming the spherical globule  into a cigar shape. Eventually the two sides of the sheet meet, creating  a seam \Revision{with an unfavorable line tension due to unsatisfied subunit contacts}.  As the seam gradually fills in, the elastic energy associated with such an acute deformation forces the complex toward a more spherical shape. As in systems with spontaneous curvature, the hexamer shells exhibit the 12 five-fold vacancy defects required by topology. If pentamers are present they eventually fill these holes (as in Fig.~\ref{fig:Fig2-WithC0-Traj} above), but for simplicity we consider systems containing only hexamers here. The large shells are roughly but not perfectly icosahedral, presumably reflecting slow defect reorganization on assembly timescales.

The size of the assembled shell is limited by the finite system size of our simulations. Importantly, the same limitation occurs within cells when the cargo undergoes phase separation into a single complex whose size is limited by the enzyme copy number (\eg the procarboxysome precursor to carboxysome assembly \cite{Cameron2013,Chen2013}). We therefore investigated the dependence of assembly morphologies on system size, as a function of the shell bending modulus, $\kshell$ (controlled by the parameter $\eangle$). Specifically, at each value of $\kshell$ we performed a series of simulations in which the maximum size of the cargo globule was controlled by changing the system size with fixed total cargo concentration and hexamer chemical potential (section~\nameref{sec:methods}). An example assembly trajectory for a small system is shown in Fig.~\ref{fig:Fig5-NoC0-Traj}C.

As shown in Fig.~\ref{fig:Fig6-kShell-Size}, we observe a minimum globule size required for complete shell assembly, which linearly increases with $\kshell$. We observe complete wrapping for all system sizes above this threshold. Below the threshold size, assembly stalls with one or more open seams remaining; examples of this configuration are shown for a low and high bending modulus in Fig.~\ref{fig:Fig6-kShell-Size}. Interestingly, while the pentameric defects are roughly equally spaced within large shells, small shells assembled with extremely low values of $\kshell$ tend to exhibit adjacent vacancy pairs (Fig.~\ref{fig:Fig5-NoC0-Traj}C, final frame). This defect morphology focuses curvature  in a region with no elastic energy (the vacancy) while reducing the number of unsatisfied hexamer edges.

\begin{table}[!ht]
\centering
\caption{
{Effect of parameters on shell size}}
\begin{tabular}{ ll}
\hline
	 Increasing parameter {\bf decreases} shell size &  \\	
\hline			
 	shell-cargo interaction* & $\varepsilon_{\mathrm{SC}}$ \\
shell-shell interaction& $\varepsilon_{\mathrm{SS}}$\\
pentamer-hexamer affinity/hexamer-hexamer affinity & $\varepsilon_{\mathrm{ph}}/\varepsilon_{\mathrm{hh}}$ \\	
pentamer/hexamer stoichiametric ratio& $\rho_{\mathrm{p}}/\rho_{\mathrm{h}}$	\\
shell subunit/cargo stoichiometric ratio& $\rhoh/\rhoc$\\
shell bending modulus (with spontaneous curvature)** & $\kshell$ \\
\noalign{\smallskip}
\hline			
Increasing parameter {\bf increases} shell size &  \\	
\hline
cargo-cargo interaction*** &$\varepsilon_{\mathrm{CC}}$\\		
shell bending modulus (with no spontaneous curvature)** & $\kshell$ \\
\\		
\end{tabular}
\begin{flushleft}
\small{*At high $\varepsilon_{\mathrm{SC}}$, over-nucleation leads to a decrease in shell size.
**Increasing $\kshell$ disfavors deviations from the shell spontaneous curvature, and thus favors small shells in the case of high spontaneous curvature or large shells in the case of low spontaneous curvature.
***Two step assembly leads to larger shells than single step pathways; however, sufficiently high values of $\ecc$ induce over-nucleation which decreases shell size.}
\end{flushleft}
\label{table1}
\end{table}

To understand these results, in section ~\nameref{S2_Thermodynamics} we present a calculation of the equilibrium shell size distribution for subunits with no spontaneous curvature and stoichiometrically limiting cargo. We restrict the ensemble to spherical shells as observed in the simulations. While the aggregates are large and polydisperse without cargo, the calculation shows that  cargo leads to a minimum free energy spherical shell size (Figs. \ref{fig:FigS8-Theory-NoC0-AvgSize} and \ref{fig:FigS9-ShellSize-Dist}). 

The linear relationship between minimum shell size and bending modulus can be understood from our equilibrium model by comparing the excess free energy difference $\dgWrap$ between the complete shell and an unwrapped globule (see section ~\nameref{S2_Thermodynamics}). For the simulated conditions, the size and shape of the cargo globule is essentially the same in each of these states, and thus the free energy difference for a globule wrapped by $\nh$ hexamers in Eq.~\ref{eq:OmegaNoPentNoCargo} simplifies to
\begin{align}
\dgWrap = 8 \pi \kshell + \dgpent + \dmu \nh
\label{eq:dgWrap}
\end{align}
with $\dmu = \gh + \ghc - \muh$, $\dgpent$ as the free energy due to the 12 pentameric vacancies, $\gh(\ess)$ as the hexamer-hexamer interactions free energy, $\ghc(\esc)$ as the hexamer-cargo free energy, 
and $\muh=\kt \log(\rhoh)$ the chemical potential of unassembled hexamers at concentration $\rhoh$. The term $8 \pi \kshell$ describes the bending energy of the complete shell. The minimum globule size $\nStar$ corresponds to the locus of parameter values at which $\dgWrap=0$, giving
\begin{align}
\nStar = \frac{8 \pi}{-\dmu} \kshell + \frac{\dgpent}{-\dmu}
\label{eq:nstar}
\end{align}
A linear fit to the simulation results for $n*$ results in $\dmu=-2.4$ and $\dgpent=80.5\kt$, or $6.7 \kt$ per pentameric defect. Plugging in $\rhoh=10^{-3}$ subunits/$\du^3$ and $\ghc=-8.1 \kt$ for $\esc=7.0$ (using the estimate from Perlmutter \etal \cite{Perlmutter2016}) then results in $\gh\approx -0.45 \kt$. This value and the fit value of $\dgpent$ are reasonably close to interactions estimated from the relationship between the shell-shell dimerization free energy $\gh$ and potential well-depth $\ess$ for a similar model in Perlmutter \etal \cite{Perlmutter2016}. Thus, the simulation results are consistent with the minimum stable shell size predicted by the theory.

\section{Conclusions}
We have used computational and theoretical modeling to investigate factors that control the assembly of a protein shell around a fluid cargo. We have focused on two limiting regimes of protein interaction geometries --- high spontaneous curvature that drives the formation of small shells, and zero spontaneous curvature that favors assembly of flat sheets or polydisperse shells. In both regimes the presence of cargo can significantly alter the size distribution of assembled shells. For high spontaneous curvature, encapsulated cargo tends to increase shell size, whereas for shell proteins with low (or zero) spontaneous curvature cargo templating provides a mechanism to drive shell curvature and thus tends to reduce shell size. These results could provide a qualitative explanation for experimental observations on different systems in which full microcompartment shells were either larger or smaller than empty shells \cite{Lassila2014,Sutter2017,Cai2016,Mayer2016,Lehman2017}.

Our simulations identify a combination of kinetic and thermodynamic mechanisms governing microcompartment size control.  At equilibrium, the shell size is determined by the stoichiometry between cargo and shell subunits, with an excess of cargo or shell protein respectively favoring larger or smaller shells. Similarly, a high surface energy (high cargo surface tension and weak shell-cargo interactions) favors larger shells whereas a strong shell bending modulus favors shells closer to the preferred size.  Although dynamical simulations exhibit similar qualitative trends to these equilibrium results, we observe significant kinetic effects as well. Fast cargo coalescence relative to rates of shell assembly favors larger shells, since closure of an assembling shell prevents further cargo aggregation. Thus, the shell size is strongly correlated to the assembly pathway, with two-step assembly leading to larger shells than single-step pathways. Although many factors likely control shell size in biological systems, this result is consistent with the observations of small empty shell assemblies \cite{Lassila2014,Sutter2017,Cai2016,Mayer2016} and the fact that $\beta$-carboxysomes (which assemble by two step pathways \cite{Cameron2013,Chen2013}) tend to be larger and more polydisperse than $\alpha$-carboxysomes (which experiments suggest assemble by one-step pathways \cite{Iancu2010,Cai2015}).

Our results for shell proteins without spontaneous curvature \Revision{build} upon Rotskoff and Geissler \cite{Rotskoff2017}, which identified a kinetic mechanism in which cargo templating drives shell curvature, and shell closure eventually arrests assembly. Their mechanism proceeds by two-step assembly, with initial nucleation of a cargo globule followed by assembly of shell subunits, but requires that rates of subunit arrival are at least 10 times faster than cargo arrival rates \cite{Rotskoff2017}. However, it is unclear how many physical microcompartment systems may fit this criteria, \Revision{ and our results suggest other mechanisms may play important roles in microcompartment assembly.
Firstly, if cargo is stoichiometrically limiting then the finite-pool mechanism can result in finite shell sizes, with the coalesced cargo still providing a template for shell curvature.} Secondly, subunits with spontaneous curvature can form complete shells even under conditions of excess cargo or fast coalescence rates that lead to large cargo aggregates (Fig.\ref{fig:Fig3-WithC0-AvgSize} D), as observed for carboxysome assembly in cells \cite{Cameron2013}. Thus, biological microcompartments with some degree of preferred shell curvature could robustly assemble over a much wider parameter space than systems without spontaneous curvature. Intriguingly, the recent atomic-resolution microcompartment structure from Sutter \etal \cite{Sutter2017} suggests that different hexamer or pseudo-hexamer species have different preferred subunit-subunit angles, and thus the spontaneous curvature may depend on the shell composition.  We will investigate this in a future work.

The importance of spontaneous curvature  to a particular BMC system could be investigated by comparing our computational predictions to experimental shell size distributions measured for varying cargo/shell protein stoichiometries and interaction strengths. While such tests would be most straightforward to perform in vitro, they could be performed in vivo by varying expression levels of various shell proteins or the enzymatic cargoes. Of particular interest would be a comparison between the shell size distribution in the presence and absence of cargo. However, note that we have focused on extreme limits (high spontaneous curvature or zero spontaneous curvature);  systems with moderate shell spontaneous curvature may exhibit less dramatic cargo effects. Also note that the effective shell spontaneous curvature depends on the stoichiometries of different shell protein species; \eg, overexpressing pentamers would shift the size distribution toward smaller shells (Fig. \ref{fig:Fig2-WithC0-Traj} D). 

These results have implications for targeting new core enzymes to BMC interiors. Recent experiments have shown that alternative cargoes can be targeted to BMC interiors by incorporating encapsulation peptides that mediate cargo-shell interactions, but that relatively small amounts of cargo were packaged \cite{Parsons2010,Choudhary2012,Lassila2014,Condezo2017}. Our previous simulations showed that assembly of full shells requires both cargo-shell and cargo-cargo (direct or mediated) interactions. Here, we see that the strength of cargo-cargo interactions can not only affect the efficiency of cargo loading, but also the size of the containing shell.

\section*{Acknowledgments}
We are grateful to Fei Cai, Cheryl Kerfeld, Grant Rotskoff, and Phill Geissler for insightful discussions, and we additionally thank Phill for incisive comments on the manuscript. This work was supported by Award Number R01GM108021 from the National Institute Of General Medical Sciences and the Brandeis Center for Bioinspired Soft Materials, an NSF MRSEC,  DMR-1420382. Computational resources were provided by NSF XSEDE computing resources (Maverick, XStream, Bridges, and Comet) and the Brandeis HPCC which is partially supported by DMR-1420382.

\bibliographystyle{unsrt}

\clearpage
\onecolumngrid
\section*{Supporting information}

\setcounter{section}{0}
\renewcommand{\thesection}{S\arabic{section}}
\setcounter{equation}{0}
\renewcommand{\theequation}{S\arabic{equation}}

\section*{S1. Model Details}
\label{S1_ModelDetails}

\Revision{Our model represents subunits as rigid bodies comprised of pseudoatoms arranged to capture the directional attractions and shape of microcompartment pentamer and hexamer oligomers.
In comparison to earlier studies with patchy spheres (e.g. \cite{Schwartz1998,Hagan2006,Wilber2007,Baschek2012}), multi-pseudoatom subunits better describe the subunit excluded volume shape \cite{Nguyen2007, Rapaport2004, Elrad2010}, which we find to be important for representing assembly around many-molecule cargoes.  See Ref. \cite{Hagan2014} for a comparison of these approaches.}

In our model, all potentials can be decomposed into pairwise interactions. Potentials involving shell subunits further decompose into pairwise interactions between their constituent building blocks -- the excluders, attractors, `Top', and `Bottom' pseudoatoms.
It is convenient to state the total energy of the system as the sum of three terms, involving shell-shell ($U\sub{SS}{}$), cargo-cargo  ($U\sub{CC}{}$), and shell-cargo ($U\sub{SC}{}$) interactions, each summed over all pairs of the appropriate type:
\begin{align}
U = & \sum_{\mathrm{shell\ }{i}} \sum_{\mathrm{shell\ }{j < i}} U\sub{SS}{}
  + \sum_{\mathrm{cargo\ }{i}} \sum_{\mathrm{cargo\ }{j<i}} U\sub{CC}{}
  + \sum_{\mathrm{shell\ }{i}} \sum_{\mathrm{cargo\ }{j}} U\sub{SC}{}
\end{align}
where $\sum_{\mathrm{shell\ }{i}} \sum_{\mathrm{sub\ }{j < i}}$ is the sum over all distinct pairs of shell subunits in the system, $\sum_{\mathrm{shell\ }{i}} \sum_{\mathrm{cargo\ }{j}}$ is the sum over all shell-cargo particle pairs, etc.

\textbf{Shell-shell interaction potentials.}
The shell-shell potential $U\sub{SS}{}$ is the sum of the attractive interactions between complementary attractors, and geometry guiding repulsive interactions between `Top' - `Top', `Bottom' - `Bottom', and `Top' - `Bottom' pairs.
 There are no interactions between members of the same rigid body. Thus, for notational clarity, we index rigid bodies and non-rigid pseudoatoms in Roman, while the pseudoatoms comprising a particular rigid body are indexed in Greek. For subunit $i$ we denote its  attractor positions as $\{\mathbf{a}_{i\alpha}\}$ with the set comprising all attractors $\alpha$, its `Top' position $\mathbf t_{i}$, `Bottom' position $\mathbf b_{i}$ and, for the case of subunits with no spontaneous curvature, the `M' pseudoatom at the center of the subunit in the plane of the attractors, as $\mathbf m_{i}$.

\newcommand{\LJ}[1]{ \mathcal{L}_{#1} }
\newcommand{\WCA}[1]{\L}
\newcommand{\Morse}[1]{ \mathcal{M}_{#1} }
\newcommand{\Coulomb}[1]{ \mathcal{C}_{#1} }

 The shell-shell interaction potential between two subunits $i$ and $j$ is then defined as:
\begin{align}
\label{Ucc}
U\sub{SS}{}(\{\mathbf a_{i\alpha}\},\mathbf t_{i}, \mathbf a_{j}, \mathbf t_{j}) & =
     \eangle \WCA{} \left(
    \left|\mathbf{t}_{i} - \mathbf{t}_{j} \right|,
    \ \sigma\sub{t}{,ij} \right)
    \nonumber \\
  &  +
     \eangle \WCA{} \left(
    \left|\mathbf{b}_{i} - \mathbf{b}_{j} \right|,
    \ \sigma\sub{b}{} \right)
    \nonumber \\
  &  +
     \eangle \WCA{} \left(
    \left|\mathbf{b}_{i} - \mathbf{t}_{j} \right|,
    \ \sigma\sub{tb}{} \right)\mathcal{I}_\text{H}(i)\mathcal{I}_{H}(j)
    \nonumber \\
  &  +
      \WCA{} \left(
    \left|\mathbf{m}_{i} - \mathbf{m}_{j} \right|,
    \ \sigma\sub{m}{} \right)\mathcal{F}_\text{H}
    \nonumber \\
   & +
    \sum_{\alpha,\beta}^{N_{\text{a}i},N_{\text{a}j}} \ess \Morse{} \left(
    \left|\mathbf{a}_{i\alpha} - \mathbf{a}_{j\beta} \right|,
    \ r\sub{0}{}, \varrho, \rcut^\text{att} \right)
    \nonumber \\
\end{align}
The function $\WCA{}$ is defined as the repulsive component of the Lennard-Jones potential shifted to zero at the interaction diameter:
\begin{align}
\WCA{}(x,\sigma) \equiv \theta(\sigma-x)
\left[
	\left(\frac{\sigma}{x}\right)^{12} -1 \right]
\label{eq:LJ}
\end{align}
with $\theta(x)$ the Heaviside function.
The function $\Morse{}$ is a Morse potential:
\begin{align}
\Morse{}(x,r\sub{0}{},\varrho,\rcut) & =
\theta(\rcut-x) \times \nonumber \\
& \left[ \left(e^{\varrho\left(1-\frac{x}{r\sub{0}{}}\right)} - 2 \right)e^{\varrho\left(1-\frac{x}{r\sub{0}{}}\right)}
 -  V_\text{shift}(\rcut) \right]
\label{eq:Morse}
\end{align}

with  $V_\text{shift}(\rcut)$ the value of the (unshifted) potential at $\rcut$.

 The parameter $\ess$ sets the strength of the shell-shell attraction at each attractor site, $N_{\text{a}i}$ is the number of attractor pseudoatoms in subunit $i$, and $\eangle$ scales the repulsive interactions that enforce the geometry. The function $\mathcal{I}_\text{H}(i)$ is 1 if subunit $i$ is a hexamer and 0 if a pentamer; thus the term $\mathcal{I}_\text{H}(i)\mathcal{I}_\text{H}(j)$ specifies that we only enforce Top-Bottom interactions between pairs of hexamers. We included this factor because we found that Top-Bottom interactions between hexamers and pentamers slow the process of pentamers filling in holes in hexamer shells (see the main text), and pentamer-pentamer interactions are irrelevant. The factor $\mathcal{F}_\text{H}=0$ for subunits with $T{=}3$ preferred curvature and $\mathcal{F}_\text{H}=1$ for subunits with zero spontaneous curvature, so that the `M' pseudoatoms are included only for the latter case. As mentioned above, the `M' pseudoatoms were only needed in the limit of small $\kshell$, which we only considered for subunits without spontaneous curvature.

\textbf{Shell-shell interaction parameter values.} \textit{Attractors:} The strength of attractive interactions is parameterized by the well-depth $\ess$ for a pair of attractors on hexamers as follows. Hexamer-Hexamer edge attractor pairs (A2-A6, A3-A5, and A5-A6) have a well-depth of $\ess$. Because vertex attractors (A1, A4) have multiple partners in an assembled structure, whereas edge attractors have only one, the well-depth for the vertex pairs (A1-A4 and A4-A4)  is set to $0.5\ess$. Similarly, for pentamer-hexamer interactions,  the well-depth for edge attractor pairs (A2-A5, A3-A6) is $\eph$, while the vertex interaction pairs (A1-A4 and A4-A4) have $0.5\eph$.
\Revision{We set the ratio $\eph/\ess$=1.3 so that simulations without cargo form  T=3 shells, or shells close in size to T=3 (see Fig.~\ref{fig:Fig4-WithC0-Pentamer}) for the parameter ranges we consider with cargo.  Note that we cannot compare exact parameter ranges with and without cargo, since we focus on conditions for which the cargo is required for shell nucleation. Therefore, we performed our empty shell simulations with higher subunit-subunit interaction strengths, $\ess=2.6$, but maintaining the ratio $\eph/\ess$=1.3.  Interestingly, complete shells at the low stoichiometric ratio $\rhop/\rhoh=0.3$ incorporated excess hexamers during assembly, but these were eventually shed resulting in complete shells with 12 pentamers and 20 hexamers. }

  \textit{Repulsive interactions:} The `Top' and `Bottom' heights, or distance out of the attractor plane, are set to $h=1/2 \rb$, with $\rb=1$ the distance between a vertex attractor and the center of the pentagon.
For simulations of shells with $T{=}3$ preferred curvature,  $\sigma\sub{tb}{}=1.8\rb$ is the diameter of the `Top' - `Bottom' interaction (this prevents subunits from binding in inverted configurations \cite{Johnston2010}), and $\sigma\sub{b}{}=1.5\rb$ is the  diameter of the `Bottom' - `Bottom' interaction.  In contrast to the latter parameters, $\sigma\sub{t}{,ij}$ the effective diameter of the `Top' - `Top' interaction, depends on the species of subunits $i$ and $j$; denoting a pentagonal or hexagonal subunit as `$\text{p}$' or `$\text{h}$' respectively, $\sigma\sub{t,pp}{}=2.1\rb$, $\sigma\sub{t,hh}{}=2.4\rb$, and $\sigma\sub{t,ph}{}=2.2\rb$. The parameter  $r\sub{0}{}$ is the minimum energy attractor distance, set to $0.2 \rb$, $\varrho=4\rb$ determines the width of the attractive interaction, and $\rcut^\text{att}=2.0\rb$ is the cutoff distance for the attractor potential. Since the interactions just described are sufficient to describe assembly of the shell subunits, we included no excluder-excluder interactions and $\mathcal{F}_\text{H}$ is zero for simulations of shells with preferred curvature.
For flat subunits, the diameter of the `Top' - `Top' interaction is equal to the diameter of  `Bottom' - `Bottom' interaction, $\sigma\sub{t}{,hh}=\sigma\sub{b}{}=2.226\rb$, $\sigma\sub{tb}{}=2.0 \rb$, and $\sigma\sub{m}{}=2.026\rb$ is the effective diameter of the  middle excluders `M'.  Attractor parameters are the same as for $T{=}3$ subunits.

\textbf{Cargo-cargo interactions.}
The interaction between cargo particles is given by
\begin{eqnarray}
\label{Ulj}
U\sub{CC}{}(\{\mathbf l_{i}\}, \{\mathbf l_{j}\})  &=&
    \sum_{i<j}^{N_{l}} \ecc \LJ{} \left(
    \left|\mathbf{l}_{i} - \mathbf{t}_{j} \right|,
    \ \sigma\sub{C}{},\rcut^\text{c} \right)
\label{eq:Ucargo}
\end{eqnarray}
with $\LJ{}$ the full Lennard-Jones interaction:
\begin{align}
\LJ{}(x,\sigma,\rcut)= & \theta(x-\rcut) \times \nonumber \\
& \left\{ 4\left[ \left(\frac{x}{\sigma}\right)^12 - \left(\frac{x}{\sigma}\right)^6 \right] -V_\text{shift}(\rcut) \right\}
\end{align}
 and $\ecc$ is an adjustable parameter which sets the strength of the cargo-cargo interaction, $N\sub{l}{}$ is the number of LJ particles, the cargo diameter is $\sigma\sub{C}{}=\rb$ and the cutoff is  $\rcut^\text{c}=3\sigma_\text{C}$.

\textbf{Shell-cargo interactions.}
The shell-cargo interaction is modeled by a short-range repulsion between cargo-excluder and cargo-`Top' pairs representing the excluded volume, plus an attractive interaction between pairs of cargo particles and hexamer `Bottom' pseudoatoms. (We do not consider pentamer-cargo attractions because there is no experimental evidence for them.) For subunit $i$ with excluder positions $\{\mathbf{x}_{i\alpha}\}$ and `Bottom' psuedoatom $\mathbf{b}_{i}$, and cargo particle $j$ with position $\mathbf R_j$, the potential is:
\begin{align}
\label{uads}
U\sub{SC}{}(\{\mathbf x_{i\alpha}\}, \mathbf R_j) &=
    \sum_{\alpha}^{N\sub{x}{}} \WCA{} \left(
    | \mathbf{x}_{i\alpha} - \mathbf R_j |,
    \sigma\sub{ex}{}\right) \\
    &+ \sum_{\alpha}^{N\sub{t}{}} \WCA{} \left(
    | \mathbf{t}_{i\alpha} - \mathbf R_j |,
    \sigma\sub{t}{}\right) \\
	&+
    \sum_{\alpha}^{N\sub{b}{}} \esc \Morse{} \left(
    \left|\mathbf{c}_{i\alpha} -  \mathbf R_j \right|,
    \ r\sub{0}{}, \varrho^\text{SC}, \rcut^\text{SC} \right)\mathcal{I}_\text{H}(i) \nonumber
\end{align}
where $\esc$ parameterizes the shell-cargo interaction strength, $N_\text{x}$, $N_\text{t}$, and $N_\text{b}$ are the numbers of excluders, `Top', and `Bottom' pseudoatoms on a shell subunit, $\sigma\sub{ex}{} =0.5 \rb$ and $\sigma\sub{t}{} =0.5 \rb$ are the effective diameters of the Excluder - cargo and `Top' - cargo repulsions, $r\sub{0}{}^\text{SC}=0.5 \rb$ is the minimum energy attractor distance, the width parameter is $\varrho^\text{SC}=2.5\rb$, and the cutoff is set to $\rcut^\text{SC}=3.0\rb$. Finally, the term $\mathcal{I}_\text{H}(i)$ specifies that only hexamers have attractive interactions with cargo.
\pagebreak
\section*{S2.~Thermodynamics}
\label{S2_Thermodynamics}

In this section we extend the equilibrium model of Perlmutter \etal \cite{Perlmutter2016} for shell assembly around a multi-molecule cargo   to allow for formation of shells with any size. A similar approach was recently considered in Rotskoff and Geissler \cite{Rotskoff2017}.

We consider shells composed of $n=\nh+\np$ subunits, with $\nh$ hexamer subunits and $\np$ pentamer subunits (or pentameric vacancies if no pentamer proteins are present). We will  assume that each shell contains the minimum number of pentamers (or pentameric vacancies) dictated by topology, $\np=12$.  Based on the fact that experiments on BMCs  and simulations \cite{Perlmutter2016,Rotskoff2017} exhibit predominantly \Revision{spherical shell geometries that are roughly but imperfectly icosahedral, we do not consider spherocylinders or other geometries\cite{Nguyen2005}), but we discuss the limits of this assumption below}. Following Lidmar, Mirny, and Nelson (LMN) \cite{Lidmar2003} and Nguyen, Bruinsma, and Gelbart (NBG) \cite{Nguyen2005}, we consider the elastic energy for icosahedral shells as a function of their radius of curvature $R$ in the continuum limit, thus assuming that an icosahedral closed shell geometry is possible for any size $R$.

Each shell encapsulates $\nc$ cargo molecules,  given by $\nc= \rhocbar n^{3/2}$, with $\rhocbar=\frac{a^3 \rhoc}{6 \sqrt{\pi}}$ and $a^2$ the area per shell subunit (measured at the inner surface of the shell), and $\rhoc$ the cargo density (which we assume is approximately its liquid density). Shells assemble from a solution of free pentamers, hexamers, and cargo molecules with concentrations $\rhop$, $\rhoh$, and $\rhoc$.

The total free energy density is given by
\begin{align}
f_\text{tot} & =  \sum_{\alpha=\text{p},\text{h},\text{c}} \kt \rho_\alpha [ \ln ( \rho_\alpha v_0 ) -1 ] + \nonumber \\
& \sum_{n=\Nmin}^{\infty} ( \kt \rho_n [\ln (\rho_n v_0) - 1]\nonumber +  \rho_n G(n)) \\
\label{eq:freeEnergyTot}
\end{align}
where the index $\alpha$ runs over free pentamers (p), hexamers (h), and cargo molecules (c), $v_0=a^3$ is a standard state volume, $\rho(n)$ is the concentration of shells with $n$ subunits, $G(n)$ is the free energy in such a shell arising from shell-shell and shell-cargo interactions, and $\Nmin$ is the minimum shell size allowed by geometry (\eg 12 pentamers).
We then minimize $f_\text{tot}$  with respect to $\{ \rho_n\}$, subject to the constraint that the total concentrations of pentamer, hexamer, and cargo molecules $\rhopT$, $\rhohT$, and $\rhocT$ are fixed:
\begin{align}
\rhopT = & \rhop + \np \sum_{n=\Nmin}^{\infty} \rho_n \nonumber \\
\rhohT = & \rhoh + \sum_{n=\Nmin}^{\infty} (n-\np) \rho_n \nonumber \\
\rhocT = & \rhoc +  \sum_{n=\Nmin}^{\infty} \rhocbar n^{3/2} \rho_n
\label{eq:massConservation}.
\end{align}
The minimization results in the law of mass action for  concentrations of shells \cite{Safran1994,Perlmutter2016}:
\begin{align}
\rho(n) = & \exp\left[ -\Omega(n)/\kt\right] \nonumber \\
\Omega(n) = & \left(G(n) - \np \mup - (n-\np) \muh - \rhocbar n^{3/2} \muc\right)
\label{eq:LMA},
\end{align}
  where $\Omega(n)$ is the excess free energy which includes the mixing entropy penalty associated with removing subunits and cargo particles from solution, with chemical potentials  $\mu_\alpha=\kt \ln \left( v_0 \rho_\alpha\right)$ for $\alpha=\{\text{p},\text{h},\text{c}\}$.

We define the interaction free energy $G(n)$ as:
\begin{align}
G(n) = & \Eelastic(n) + \dgpent +  \nonumber \\
& (n-\np) (\gh+\ghc) + \surfTension a^2 n + \rhocbar n^{3/2} \mucliq \
\label{eq:Gn},
\end{align}
with $\dgpent = \np (\gp +\gpc)$ (provided pentamers are present) and $\gp$ and $\gh$ as the shell shell-shell binding free energy per pentamer or hexamer (we assume the shells are large enough that there are no direct pentamer-pentamer interactions), $\gpc$ and $\ghc$ the shell-cargo interaction free energy strengths,  $\mucliq$ the chemical potential of the cargo subunits within the packaged globule, and $\surfTension$ the surface tension of the cargo globule. If there are no pentamers present, then $\dgpent$ accounts for the 12 pentameric vacancies.

The term $\Eelastic$ gives the elastic energy of the shell arising from bending and stretching deformations, including the contributions of the 12 disclinations required by topology. In the case of a fluid membrane, the bending energy is given by the ratio of its curvature radius $R$ to its spontaneous curvature $\Rsp$ by the Helfrich energy, $\Ehelfrich(R/\Rsp)$, with  \cite{Helfrich1973}
\Revision{
\begin{align}
\Ehelfrich(m) = 8 \pi \kappa\left(1 -  2m + m^2\right) + 4 \pi \kappa_\text{G}
\label{eq:Helfrich},
\end{align}
}
with $\kappa$ and $\kappa_\text{G}$ the mean and Gaussian curvature moduli.

The deformation energy for an elastic icosahedral shell without 
 spontaneous curvature was derived by LMN \cite{Lidmar2003} and then \Revision{approximately} extended to include spontaneous curvature by NBG \cite{Nguyen2005}. The behavior depends on the dimensionless F\"{o}ppl-von K\'{a}rm\'{a}n number (FvK), $\gamma=Y R^2/\kshell$ with $Y$ the 2D Youngs modulus,   which  gives the relative importance of bending and stretching. Stretching energy dominates over bending when  $\gamma > \gammab\approx130$, driving buckling of the shell  \cite{Lidmar2003,Nguyen2005}. Below the buckling threshold, the elastic energy is given by
\begin{align}
\Eelastic(\gamma,\gammaSp) \approx & 6 \kappa B \gamma/\gammab + \Ehelfrich(\sqrt{\gamma/\gammaSp}) \qquad \mbox{for }   \gamma<\gammab 
\label{eq:Eelastic}.
\end{align}
where $\gammaSp=Y \Rsp^2/\kshell$ is the FvK for a shell at its minimum energy size ($R=\Rsp$), and the first term gives the energy arising from the
\Revision{elastic interactions between the 12 disclinations for an icosahedral structure, with $B\approx\pi/3$ a numerical constant \cite{Bowick2000,Lidmar2003}. The elastic energy from the defect interactions grows quadratically with shell size, until $\gammab$  when it becomes favorable to screen the interaction by buckling.} Above this threshold, the elastic energy in the absence of spontaneous curvature is given by \cite{Lidmar2003}
\begin{align}
\Eelastic(\gamma,\gammaSp=\infty) \approx & 6 \kappa B \left[ 1 + \ln(\gamma/\gammab)\right] + \Ehelfrich(0)  \mbox{ for }   \gamma>\gammab
\label{eq:EelasticBuckle}.
\end{align}
We omit the (lengthy) expression for the case of non-zero spontaneous curvature above buckling \cite{Nguyen2005}, since in the present paper we focus on the sub-buckling case for  simulations with spontaneous curvature. We will consider buckling of shells with spontaneous curvature in a future work. 

\textbf{Mean shell size.}
The mean shell size can be obtained from Eqs.~\eqref{eq:massConservation} and \eqref{eq:LMA} as a function of the chemical potentials $\mup$, $\muh$, $\muc$ using
\begin{align}
\langle n \rangle = \int n \rho(n) / \int \rho(n)
\label{eq:Shellsize}.
\end{align}
Alternatively,  since the total concentrations of each species $\rhopT$, $\rhohT$, and $\rhocT$ are the usual experimental control variables, it is convenient to numerically solve for the three unknown chemical potentials  at fixed total concentrations.

\textbf{Shell size distribution for subunits with no spontaneous curvature, in the limit of excess of hexamers}. \Revision{In this section we calculate the shell size distribution corresponding to the simulation results on hexamer subunits without spontaneous curvature. Based on the simulation results, we restrict the calculation to spherical cargo globules and icosahedral shells, so the complete shell contains $\nh$ hexamers and 12 pentameric vacancies. We discuss the limits of this restriction below.}

In the absence of cargo, and below the buckling threshold, the excess free energy in Eq.~\ref{eq:LMA} is given by
\begin{align}
\Omega(\nh) = & G_0  + \dmuprime
\label{eq:OmegaNoPentNoCargo}
\end{align}
with $G_0=8 \pi \kappa  + \dgpent$ and $\dmuprime=\dmu+6 \kappa B/\nb$ with $\nb=4 \pi \gammab \kshell/Y a^2$ the threshold buckling size.
Eq.~\eqref{eq:OmegaNoPentNoCargo} has the same form as the free energy of a system of fluid vesicles \cite{Helfrich1986} (for simplicity we are neglecting the renormalization of $\kappa$ with shell size). However, allowing for equilibrium between assembled shells and free subunits gives the form of a cylindrical micelle \cite{Safran1994}, with an exponential shell size distribution
\begin{align}
P(\nh) = \exp\left(\nh/\langle \nh \rangle \right)
\label{eq:Pn}
\end{align}
with $\langle \nh \rangle\approx \sqrt{\rhohT e^{\beta G_0}}$. Thus shells are polydisperse, with the standard deviation of shell sizes equal to the mean. Significant assembly requires a total subunit concentration exceeding the `critical concentration' \cite{Schoot2007,Hagan2006}
\begin{align}
\rhoStar \approx e^{\beta \left(\gh+6 \kappa B/\nb \right)}
\label{eq:rhoStar}.
\end{align}

As pointed out in NBG \cite{Nguyen2005}, above the buckling threshold the free energy is unstable due to the presence of the log term in  Eq.~\eqref{eq:EelasticBuckle}, and the distribution is thus highly polydisperse.

In the presence of cargo, the excess free energy is given by
\begin{align}
\Omega(\nh) = G_0  + \dmuprime \nh + \rhocbar \nh^{3/2} \dmuc
\label{eq:OmegaNoPent}
\end{align}
with the chemical potential difference now including shell-cargo interactions,
$\dmuprime=\gh +\ghc - \muh +\surfTension a^{2} + 6 \kappa B/\nb $, and the cargo chemical potential difference $\dmuc=\mucliq - \muc$.

\Revision{Under conditions of limiting cargo, the system will equilibrate at concentrations of free shell subunits and cargo such that $\dmuprime<0$ and $\dmuc>0$, with $\muc=\kt \log(\rhoc v_0)$ and $\rhoc = \rhocT - \sum_{\nh=\Nmin}^{\infty} \rhocbar \nh^{3/2} \rho_{\nh}$ accounting for the `finite-pool' of cargo particles \cite{Mohapatra2016}. The finite pool effect gives rise to a minimum in Eq.~\eqref{eq:OmegaNoPent} , and correspondingly a maximum in the shell size distribution (~\ref{fig:FigS9-ShellSize-Dist}). Note that in the thermodynamic limit, the condition $\dmuprime<0$
should make the system unstable to other structures  with a larger surface-to-area ratio, such as spherocylinders. Indeed, Cameron \etal \cite{Cameron2013} observed elongated structures when pentamer proteins were knocked out and RuBisCO was overexpressed. We do not allow for these in the present calculation because we do not observe them in our simulations, either because the system size is not large enough or because the initial coalescence of cargo into a spherical droplet makes these geometries kinetically inaccessible.}

\paragraph*{ Determination of parameter values}
\label{sec:paramValues}
Comparing predictions of the equilibrium theory against BD simulation results requires a mapping between the  interaction parameters of the theory ($\gh$, $\ghc$, $\gp$, $\gpc$, $\mucliq$, $\surfTension$, and $\rhoc$) and simulations ($\ess$, $\eph$, $\esc$, $\ecc$). For this purpose, we use the mappings estimated in \cite{Perlmutter2016}. Note that these mappings are approximate, and we have not updated them for changes in  $\eangle$ (and in the case of flat subunits, the preferred subunit-subunit angle) between the two studies. Moreover, the estimates for subunit-subunit binding affinities ($\gh(\ess)$ and $\gp(\eph)$) are calculated for subunit dimerization reactions, and thus do not fully account for differences in the translational and rotational entropy of subunits within a complete shell compared to an a dimer. Thus, we can only qualitatively compare the equilibrium theory against the simulation results. However, the fitting parameters independently estimated for subunit-subunit interactions from our measurements of the shell bending modulus described next and in Fig.~\ref{fig:Fig6-kShell-Size} agree reasonably well with the calculations from Ref. \cite{Perlmutter2016}.

\textbf{Estimating the shell bending modulus, $\kshell$.} We tune the bending modulus in our computational model by varying the parameter $\eangle$. However the angular dependence of the subunit-subunit interaction arises from a combination of nonlinear repulsive and attractive potentials, and has sufficient complexity that we could not directly calculate the bending modulus. We therefore obtained rough empirical estimates of the relationship $\kshell(\eangle)$ by measuring the change in the average energy of assembled shells as a function of $\eangle$ and/or shell size. Note that the dependence of $\kshell$ on $\eangle)$ differs for the two versions of the model (with and without spontaneous curvature).

For shells with $T{=}3$ spontaneous curvature, we extracted a complete shell containing 98 hexamers and 12 pentamers, along with cargo, which had assembled in a simulation with parameters $\ehh$=1.8, $\esc$=9.0 and $\eangle$=1. We then performed a set of BD simulations, each at a different value of $\eangle$ but with other parameters fixed, using the complete shell configuration as initial conditions. In each simulation we performed $10^{5}$ time steps to allow relaxation under the new value of $\eangle$, followed by an additional $5 \times 10^4$ time steps during which we measured the total energy of the shell, $\Gshell(\eangle)$, including all shell-shell attractive and repulsive interactions (but not shell-cargo interactions since these were nearly independent of $\eangle$). We then performed two regression analyses to fit the measured dependence of $\Gshell$ on $\eangle$ according to
\begin{align}
\Gshell(\eangle) = & U_{0} + \Ebend(\eangle) \nonumber \\
\Ebend(\eangle) = & C_1 \eangle^{p} +C_2 \qquad
\label{eq:GnShell}
\end{align}
with $p=1$ (linear regression) or $p=1/2$ and $C_1$ and $C_2$ fit parameters.

The constant $U_0$ estimates the shell energy at $\eangle=0$ and thus can be interpreted as the contribution from the attractive interactions and pentamers in their unperturbed configurations, $U_{0} =\np (\gp ) + (n-\np) (\gh)$. The remainder of  Eq.~\eqref{eq:GnShell} captures the variation of shell energy with $\eangle$, and thus can be interpreted as the bending energy arising from deviations from the shell spontaneous curvature. ~\ref{fig:FigS10-Eangle-fit} shows the fits of Eq.~\eqref{eq:GnShell} to the simulation data. 

We then estimate the bending modulus from $\Ebend$ according to
\begin{align}
\Ebend(\eangle)=8\pi \kshell\left(1-\frac{n}{n_0}\right)^2
\label{eq:EelasticEangle}
\end{align}
with $n=110$ subunits in the simulated shell, and $n_0=32$ the number of subunits in a shell with radius equal to its spontaneous curvature $\Rsp$. For $\eangle$=0.5 nonlinear and linear fits result in $\kshell$=14.5 and $\kshell$=6.2 respectively. Discriminating between these fits (or any other value of $p$) is challenging because they primarily differ near $\eangle=0$ where we are unable to obtain simulation results. Moreover, the calculated $\kshell$ depends on the value obtained for $U_{0}$. Thus, we set $\kshell=10\pm5 \kt$ as an approximate average between the two fits.

Our simulations of flat subunits explore a wider range of $\eangle$ and shell sizes than those of simulations with spontaneous curvature. Consequently, we observed more significant nonlinear effects, and a higher-order dependence of elastic energy on shell size than accounted for in Eq.~\eqref{eq:Eelastic}. Note that these nonlinearities are not consistent with the expected renormalization of bending modulus with shell size \cite{Helfrich1986}, but rather arise from the very large deviations from the preferred curvature $\Rsp=\infty$ for the small shells considered. Therefore, for each shell size considered in Fig.~\ref{fig:Fig6-kShell-Size}, we measured the interaction energy $\Gshell$ as a function of $\eangle$ following the procedure described above, and then estimated an effective value of $\kshell$ from Eqs.~\eqref{eq:GnShell} and \eqref{eq:Eelastic} with $n/n_0=0$.

\Revision{Our estimated bending modulus values are comparable to mechanical properties of carboxysomes measured by AFM. Using AFM nanoindentation experiments on $\beta$-carboxysomes, Faulkner \etal \cite{Faulkner2017} estimated a 3D Young's modulus of $E=0.6$ MPa from a linear fit or $E=80$ MPa from a Hertzian fit to the nanoindention profiles. These estimates lie below the range of Young's modulus values measured for viruses by nanoindention, $E\in[100\mbox{MPa},2\mbox {GPa}]$ \cite{Michel2006,Roos2007,May2011,Schaap2012}, thus suggesting that the carboxysome bending modulus lies below the range of corresponding bending modulus values for viruses, $\kappa\in[30,600]\kt$. }

\Revision{A lower bound on the bending modulus can be estimated from the linear fit according to thin shell elasticity as \cite{May2011a}
\begin{align}
\kshell = \frac{E h^3}{12 \left(1-\nu\right)^2}
\label{eq:KappaFromNano}
\end{align}
with $h$ the thickness of the carboxysome shell and $\nu$ the Poisson's ratio. Using $h\approx4.5$ nm estimated from the carboxysome structure \cite{Faulkner2017} and the typical Poisson's ratio for proteins $\nu=0.3$  \cite{May2011a} results in $\kshell=1.9\kt$. This is a crude estimate since the nanoindention profile is better fit by the nonlinear Hertzian model and the effective thickness $h$ typically corresponds to the minimum thickness of the shell rather than its mean thickness; for instance the effective thickness of virus shells has been estimated at $h\approx2$  \cite{May2011a}. However, from a direct comparison of the estimated Young's modulus values for carboxysomes and viruses, it is reasonable to estimate that the carboxysome bending modulus falls in the range $\kshell \in[1,25]\kt$.}
\clearpage

\setcounter{figure}{0}
\renewcommand{\thefigure}{S\arabic{figure}}

\begin{figure}[ht!]
\centering{\includegraphics[width=0.4\columnwidth]{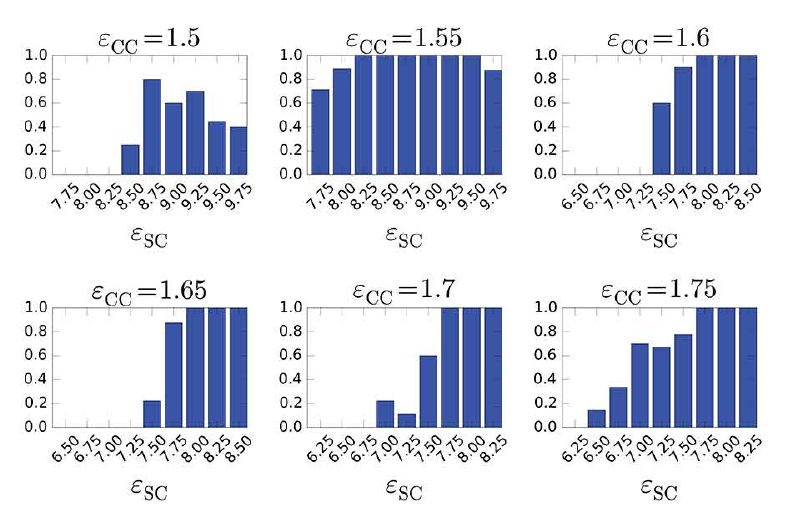}}
\caption{
{  Fraction of Brownian dynamics trials at each parameter set that lead to at least one complete shell}. A complete shell is defined as a structure in which all pentamers and hexamers have respectively five and six interactions with neighbors. Results are shown as a function of $\esc$ at indicated values of $\ecc$. Other parameters are $\ess=1.8$, $\ephr=1.3$, $\rhop/\rhoh=0.5$, and $\kshell=10\kt$.
}
\label{fig:FigS1-Yield}
\end{figure}

\begin{figure}[ht!]
\centering{\includegraphics[width=0.4\columnwidth]{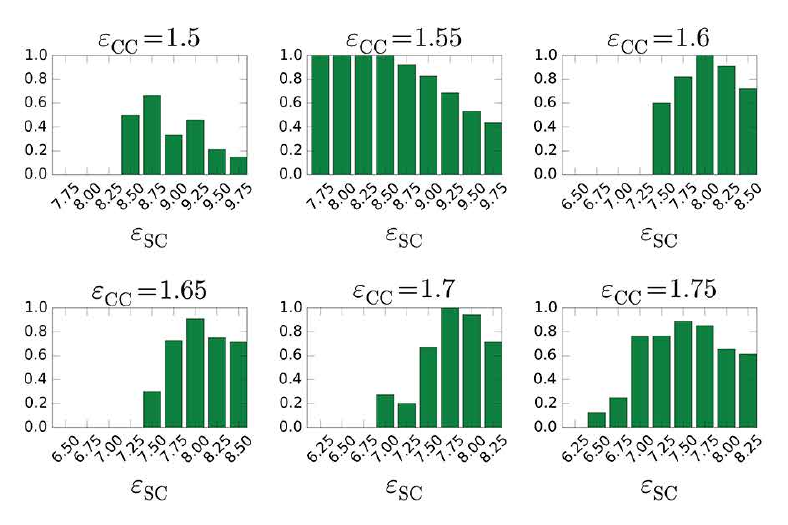}}
\caption{
{ Quality of shells}. Ratio of complete shells to the total number of shells with at least 32 subunits as a function of $\esc$ at indicated values of $\ecc$. Other parameters are as in \ref{fig:FigS1-Yield}.
}
\label{fig:FigS2-Quality}
\end{figure}

\begin{figure}[ht!]
\centering{\includegraphics[width=0.38\columnwidth]{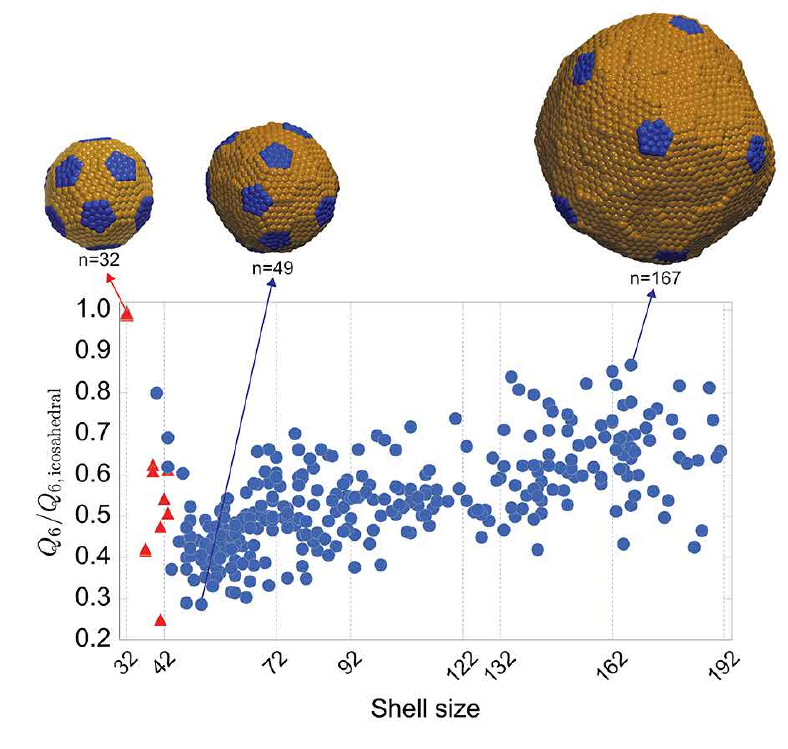}}
\caption{
{ The degree of icosahedral symmetry increases with shell size for full shells.} The bond order parameter $Q_6$ of Ref.~\cite{steinhardt1983} is shown as a function of the number of subunits in a shell, \Revision{ with
$Q_{l}=\left[ \frac{4\pi}{2l+1} \sum_{m=-l}^{l} |\bar{Q}_{lm}|^{2} \right]^{1/2} , \bar{Q}_{lm} \equiv \left< Q_{lm}(\bf{r}) \right> $%
 where the average is taken over all the geometric center of each pentamer $\bf{r}$, and  $Q_{lm}(\bf{r})$ is the ($lm\mathrm{th}$) spherical harmonic of  $\bf{r}$. Results are normalized by the value for perfect icosahedral symmetry, $Q_6=0.663$, and  \textcolor{blue}{\large{$\bullet$}} symbols correspond to the complete shells from the simulations used for Fig. \ref{fig:Fig3-WithC0-AvgSize}, while \textcolor{red}{$\blacktriangle$} symbols correspond to empty shells.}
}
\label{fig:FigS3-Symmetry}
\end{figure}

\begin{figure}[ht!]
\centering{\includegraphics[width=0.5\columnwidth]{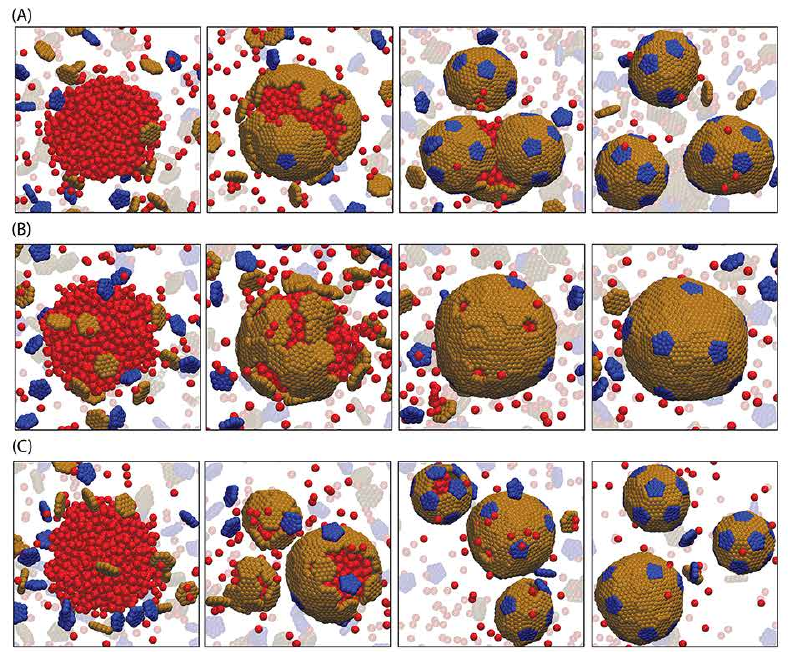}}
\caption{
{ Snapshots of assembly around a pre-equilibrated cargo globule.} These snapshots are from Brownian dynamics simulations that used an alternative initial condition (described in the text), in which cargo particles were allowed to equilibrate before introduction of shell subunits. \textbf{(A)} With $\esc$=8.0, $\ess$=2.0, $\rhop/\rhoh=0.6$, $\ephr=1.5$, and $\kshell=16\kt$ , small shells assemble and bud from the globule. At this moderate shell-cargo affinity, pentamers rapidly associate with adsorbed hexamers, driving high shell curvature. The final shells have 44-63 subunits, encapsulating 133-274 cargo particles. \textbf{(B)} With stronger shell-cargo interactions ($\esc=10$, other parameters as in (A)), hexamers adsorb rapidly and exclude pentamers from the globule. Eventually there are 12 vacancies in the hexamer lattice that are filled by pentamers. The final shell has 104 subunits encapsulating 641 cargo particles. \textbf{(C)} Further increasing the shell cargo interaction ($\esc=12$, other parameters as in (A)) leads to multiple nucleation events and polydisperse shell.  The simulation results in four complete shells containing 37-92 subunits and 116-532 cargo particles.
}
\label{fig:FigS4-PreEq-Hmgns-Traj}
\end{figure}

\begin{figure}
\centering{\includegraphics[width=.5\columnwidth]{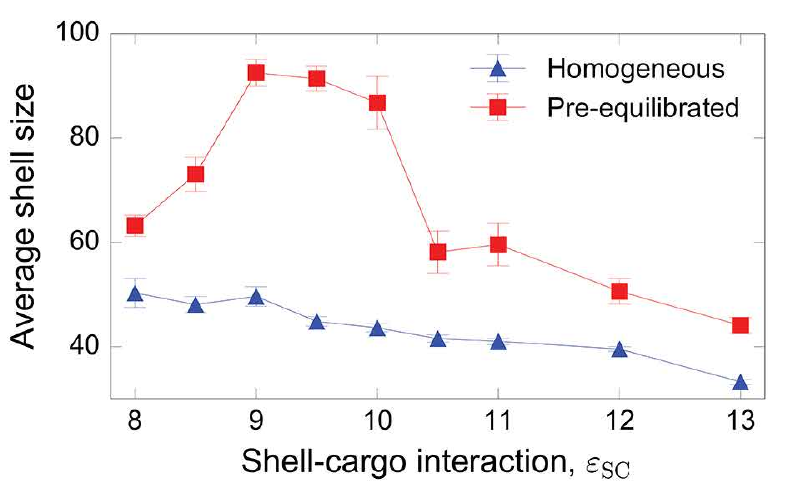}}
\caption{
{ Comparison of the mean shell size for BD simulations started from the homogeneous initial condition and pre-equilibrated globule initial conditions  for varying $\esc$.} Other parameters are $\ecc=1.5$, $\ess=2.0$, $\ephr=1.5$, $\rhop/\rhoh=0.5$, and $\eangle=1.0$ ($\kshell \approx 16 \kt$).
}
\label{fig:FigS5-PreEq-Hmgns-AvgSize}
\end{figure}

\begin{figure}[ht!]
\centering{\includegraphics[width=.85\columnwidth]{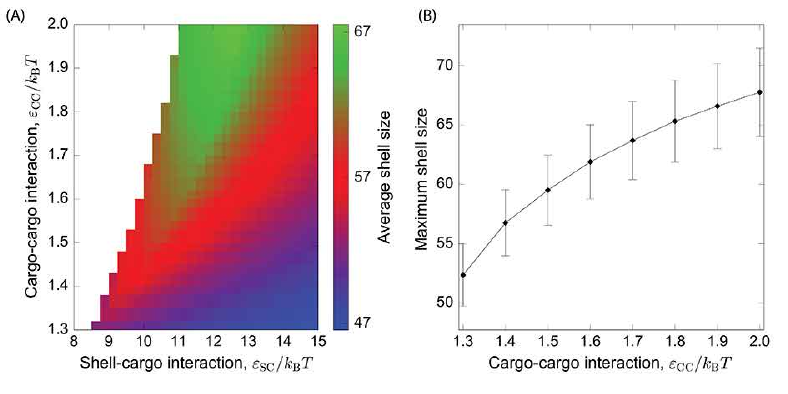}}
\caption{\textbf{(A)}{ Predictions from the equilibrium model (Eqs.  ~\eqref{eq:freeEnergyTot}--~\eqref{eq:LMA} and ~\eqref{eq:Shellsize}) for the mean shell size as a function of the cargo-cargo and shell-cargo affinities.} \textbf{(A)} Results are shown for parameters at which at least 1\% of subunits are in shells, for $\ess=1.8$, and shell bending modulus $\kshell=10\kt$. Cargo and shell volume fractions are the same as in Fig.~\ref{fig:Fig3-WithC0-AvgSize}. \textbf{(B)}  Mean and standard deviation of the equilibrium shell size distribution  as a function of cargo-cargo affinity, maximized over shell-cargo affinity. Other parameters are as in (A).
}
\label{fig:FigS6-Theory-WithC0-AvgSize}
\end{figure}
\begin{figure}[ht!]

\centering{\includegraphics[width=.5\columnwidth]{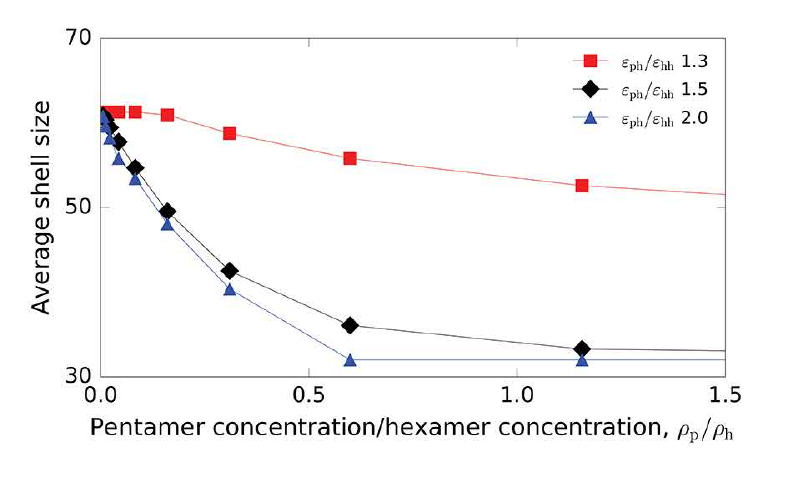}}

\caption{
{ Mean shell size predicted by the equilibrium theory (Eqs.  ~\eqref{eq:freeEnergyTot}--~\eqref{eq:LMA} and \eqref{eq:Shellsize}) as a function of pentamer/hexamer stoichiometry ratio $\rhop/\rhoh$ and pentamer/hexamer affinity ratio $\eph/\ehh$.} The theory parameters are calculated to approximately match the simulation parameters in Fig.~\ref{fig:Fig4-WithC0-Pentamer} (see section~\nameref{S2_Thermodynamics}), with $\ess=1.8$, $\kshell=10$, $ \ecc=1.65$, and  $\esc=10.0$.
}
\label{fig:FigS7-WithC0-Pentamer-THeory}
\end{figure}


\begin{figure}[ht!]
\centering{\includegraphics[width=.3\columnwidth]{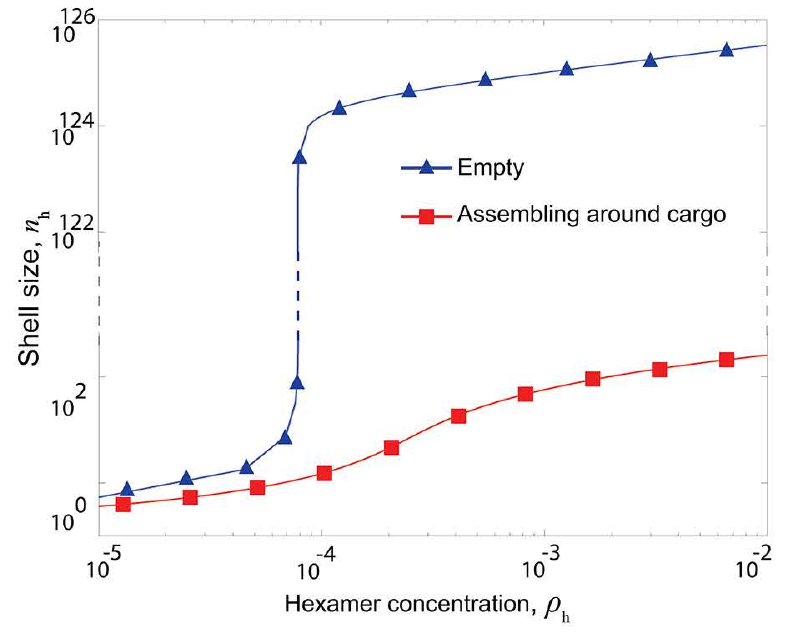}}
\caption{
{Equilibrium theory prediction of mean shell size for subunits with no spontaneous curvature restricted to icosahedral shells, in the presence (red circles) and absence (blue squares) of cargo.} The mean shell size is shown as a function of hexamer concentration, calculated from Eqs.~\ref{eq:OmegaNoPent} and \ref{eq:Shellsize} with hexamer-cargo affinity $\ghc=-8.1$ (corresponding to $\esc=7.0$, see Ref.~\cite{Perlmutter2016}), and $\kshell=20\kt$. The hexamer-hexamer affinity $\gh=-0.45$ and the energy of 12 pentameric vacancies $\dgpent=80.5\kt$ were obtained from the fit to the simulations in Fig.~\ref{fig:Fig6-kShell-Size}.
}
\label{fig:FigS8-Theory-NoC0-AvgSize}
\end{figure}


\begin{figure}[ht!]
\centering{\includegraphics[width=0.7\columnwidth]{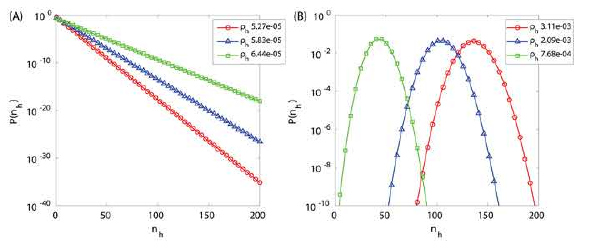}}
\caption{
{ Equilibrium shell size distribution for subunits with no spontaneous curvature.} (A) Empty shells and (B) With cargo, under conditions of excess shell subunits (limiting cargo). Size distributions are obtained by solving Eq.~\eqref{eq:OmegaNoPent}, with $\dmuc=0.18$,  $\dgpent=80$, and $\kshell=20\kt$. Other parameters are from the calculations in Ref.~\cite{Perlmutter2016} for $\esc=7.0$, $\ecc=1.7$,  and $\ehh=1.8$.
}
\label{fig:FigS9-ShellSize-Dist}
\end{figure}

\begin{figure}
\centering{\includegraphics[width=.3\columnwidth]{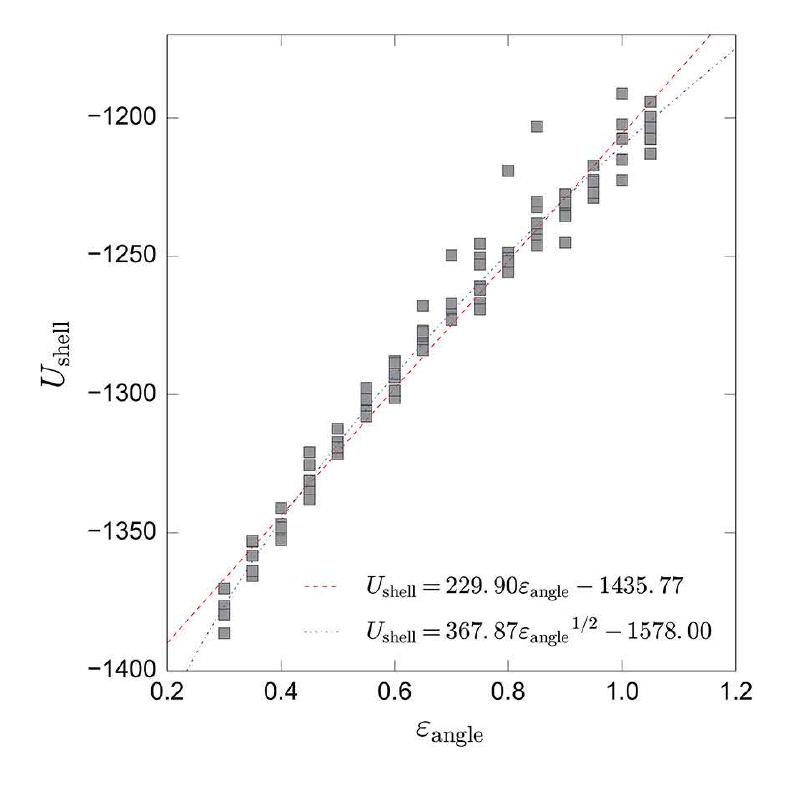}}
\caption{
{ Total interaction energy of a complete shell with preferred $T{=}3$ curvature, measured in BD simulations with different values of  $\eangle$.} The shell has 98 hexamers and 12 pentamers, and other parameters are $\ehh$=1.8, $\esc$=9.0, and $\ecc$=1.5.
}
\label{fig:FigS10-Eangle-fit}
\end{figure}


%
%
%

\end{document}